\documentclass[11pt,a4paper]{report} 

\usepackage[utf8]{inputenc}

\usepackage{eurosym}

\usepackage{url}

\usepackage[square,numbers]{natbib}
\setcitestyle{numbers}
\usepackage{fontawesome5}
\usepackage{colortbl}
\usepackage{multirow}
\usepackage{lipsum}
\usepackage{amsthm}
\usepackage{mathtools}
\usepackage{xfrac}
\usepackage[export]{adjustbox}
\usepackage{longtable}
\usepackage{makecell}
\usepackage{booktabs}
\usepackage[table,xcdraw]{xcolor}
\usepackage{amssymb}
\usepackage{tabularx}  
\usepackage{textcomp}  

\definecolor{unitree_primary}{HTML}{4C9A99}  
\definecolor{unitree_secondary}{HTML}{307FE2}  
\definecolor{unitree_accent}{HTML}{1D8348}  
\definecolor{unitree_dark}{HTML}{3F4444}  
\definecolor{unitree_light}{HTML}{F5F5F5}  
\definecolor{unitree_purple}{HTML}{8A4FFF}  
\definecolor{unitree_red}{HTML}{E74C3C}  
\definecolor{unitree_orange}{HTML}{E67E22}  
\definecolor{unitree_green}{HTML}{27AE60}  

\let\oldtexttt\texttt
\renewcommand{\texttt}[1]{\textcolor{unitree_primary}{\oldtexttt{#1}}}

\usepackage{graphicx}
\usepackage{subcaption}
\usepackage{pdflscape}
\usepackage{lmodern}
\usepackage{verbatim}
\usepackage{placeins}
\usepackage{mdframed}
\usepackage{hyperref}
\hypersetup{
    colorlinks=true,
    urlcolor=unitree_secondary,
    linkcolor=unitree_secondary,
    filecolor=unitree_accent,
    citecolor=unitree_secondary,
    breaklinks=true,  
    pdfborder={0 0 0},  
}
\usepackage{fancyvrb}
\usepackage{fixltx2e}
\usepackage{bera}
\usepackage{pdfpages}

\usepackage{pgf-umlsd}
\usepackage{tikz} 
\usetikzlibrary{shapes,arrows,positioning,calc,patterns,decorations.pathreplacing,fit}
\usepackage[draft=true]{minted}

\usepackage{fancyhdr}
\renewcommand{\headrulewidth}{0.4pt}
\renewcommand{\footrulewidth}{0.4pt}
\renewcommand{\headrule}{\hbox to\headwidth{\color{unitree_primary}\leaders\hrule height \headrulewidth\hfill}}
\renewcommand{\footrule}{\hbox to\headwidth{\color{unitree_accent}\leaders\hrule height \footrulewidth\hfill}}

\usepackage{dirtree} 
\usepackage{forest} 
\usepackage{wrapfig} 

\usepackage{algorithm} 
\usepackage{program} 

\usepackage{tocloft}

\usepackage{listings}
\lstdefinestyle{json}{
    basicstyle=\footnotesize\ttfamily,
    numbers=left,
    numberstyle=\tiny\color{gray},
    stepnumber=1,
    numbersep=10pt,
    frame=single,
    breaklines=true,
    breakatwhitespace=false,
    tabsize=2,
    keepspaces=true,
    showspaces=false,
    showstringspaces=false,
    showtabs=false,
    captionpos=b,
    morestring=[b]",
    morestring=[d]',
    comment=[l]{//},
    morecomment=[s]{/*}{*/},
    commentstyle=\color{gray}\ttfamily,
    stringstyle=\color{unitree_purple}\ttfamily,
    keywordstyle=\color{unitree_secondary}\bfseries,
}

\lstdefinelanguage{C++}{
    keywords={typedef, class, struct, enum, public, private, protected, virtual, override, const, static, inline, template, typename, using, namespace, return, if, else, while, for, do, switch, case, default, break, continue, goto, try, catch, throw, new, delete, this, true, false, nullptr, void, int, float, double, char, bool, unsigned, signed, long, short, size_t, uint32_t, uint64_t, int32_t, int64_t},
    morecomment=[l]{//},
    morecomment=[s]{/*}{*/},
    morestring=[b]",
    morestring=[b]',
    sensitive=true,
    basicstyle=\footnotesize\ttfamily,
    keywordstyle=\color{unitree_secondary}\bfseries,
    commentstyle=\color{gray}\ttfamily,
    stringstyle=\color{unitree_purple}\ttfamily,
}

\lstdefinelanguage{Python}{
    keywords={def, class, if, else, elif, while, for, return, import, from, as, try, except, finally, with, pass, break, continue, global, nonlocal, lambda, yield, assert, del, in, is, not, and, or, None, True, False, self},
    morecomment=[l]{\#},
    morecomment=[s]{"""}{"""},
    morecomment=[s]{'''}{'''},
    morestring=[b]",
    morestring=[b]',
    sensitive=true,
    basicstyle=\footnotesize\ttfamily,
    keywordstyle=\color{unitree_secondary}\bfseries,
    commentstyle=\color{gray}\ttfamily,
    stringstyle=\color{unitree_purple}\ttfamily,
}

\tikzstyle{process} = [rectangle, draw=unitree_primary, fill=unitree_primary!10, text width=2.5cm, text centered, minimum height=0.8cm, rounded corners=3pt]
\tikzstyle{arrow} = [->, >=stealth, thick, unitree_primary]
\tikzstyle{label} = [fill=unitree_light, text=unitree_dark, font=\scriptsize]
\tikzstyle{critical} = [draw=unitree_red, fill=unitree_red!10, very thick]
\tikzstyle{success} = [draw=unitree_green, fill=unitree_green!10]
\tikzstyle{warning} = [draw=unitree_orange, fill=unitree_orange!10]

\newenvironment{securityfinding}[1]{%
    \begin{mdframed}[
        backgroundcolor=unitree_red!5,
        linecolor=unitree_red,
        linewidth=2pt,
        roundcorner=5pt,
        frametitle={\textbf{Security Finding: #1}},
        frametitlebackgroundcolor=unitree_red!20,
        frametitlerule=true
    ]%
}{%
    \end{mdframed}%
}

\title{
    \vspace{-2cm}
    \color{unitree_primary}
    \textbf{\Huge The Cybersecurity of a Humanoid Robot} \\
    \vspace{0.5cm}
    \Large \color{unitree_dark} An Early Study on the Cybersecurity of Humanoid Robots via the Unitree G1 \\
    \vspace{0.3cm}
    \large \color{unitree_secondary} Technical Report
}

\author{
    \textbf{Víctor Mayoral-Vilches} \\
    \texttt{\small \faIcon{envelope} victor@aliasrobotics.com} \\
    \vspace{0.1cm}
    \textit{Alias Robotics} \\
    \vspace{0.2cm}
}

\begin{document}
\maketitle
\vspace{-1em}

\begin{abstract}
The rapid advancement of humanoid robotics presents unprecedented cybersecurity challenges that existing theoretical frameworks fail to adequately address. This report presents a comprehensive security assessment of a production humanoid robot platform, bridging the gap between abstract security models and operational vulnerabilities. Through systematic static analysis, runtime observation, and cryptographic examination, we uncovered a complex security landscape characterized by both sophisticated defensive mechanisms and critical vulnerabilities. 

Our findings reveal a dual-layer proprietary encryption system (designated ``FMX'') that, while innovative in design, suffers from fundamental implementation flaws including the use of static cryptographic keys that enable offline configuration decryption. More significantly, we documented persistent telemetry connections transmitting detailed robot state information---including audio, visual, spatial, and actuator data---to external servers without explicit user consent or notification mechanisms. These discoveries validate theoretical predictions about cross-layer vulnerability propagation while exposing implementation-specific weaknesses that could not be anticipated through modeling alone.

The assessment demonstrates that current humanoid platforms, despite employing defense-in-depth principles and hierarchical service architectures, remain vulnerable to both traditional cyber attacks and novel robot-specific exploitation vectors. Beyond documenting passive surveillance risk, we operationalized a Cybersecurity AI agent on the Unitree G1 to map and prepare exploitation of its manufacturer's cloud infrastructure, illustrating how a compromised humanoid can escalate from covert data collection to active counter-offensive operations. We argue that securing humanoid robots requires a paradigm shift toward Cybersecurity AI (CAI) frameworks that can adapt to the unique challenges of physical-cyber convergence. This work contributes empirical evidence essential for developing robust security standards as humanoid robots transition from research curiosities to operational systems in critical domains.
\end{abstract}

\tableofcontents
\newpage

\chapter{Introduction}
\label{chap:introduction}

\section{The State of Robot Cybersecurity}

To understand the cybersecurity challenges in robotics, we must first understand their fundamental architecture. Robots are networks of networks, with sensors capturing data, passing to compute technologies, and then on to actuators and back again in a deterministic manner \cite{mayoral2025offensive}. These networks can be understood as the nervous system of the robot, passing across compute \emph{Nodes} that represent neurons. Like the human nervous system, real-time information across all these computational Nodes is fundamental for the robot to behave coherently. ``Robot brains'' are built with this same philosophy. Behaviors take the form of computational graphs, with data flowing between Nodes operating intra-process, inter-process and across physical networks (communication buses), while mapping to underlying sensors, compute technologies and actuators.

The Robot Operating System (ROS) \cite{quigley2009ros} is a robotics framework for robot application development that exemplifies this architecture. ROS enables robotics developers to build these computational graphs and create robot behaviors by providing libraries, a communication infrastructure, drivers and tools to put it all together. It provides an open source codebase with a commercially friendly license that helps roboticists reduce the effort to bring up robot behaviors \cite{mayoral2025offensive}. This architectural philosophy, while enabling rapid development and modular design, also introduces security considerations at every layer of the computational graph.

The security challenges facing humanoid robots extend far beyond traditional computing systems. As previously discussed at \cite{mayoral2025offensive}, there exists a deep connection between robotic architecture and cybersecurity in robotics---more intimate than initially perceived. Robotic architecture focuses on creation, on shaping behaviors and possibilities, while cybersecurity in robotics is oriented towards offense or protection, allowing what is built to be defended. This synergy creates a unique balance when cybersecurity is approached from a knowledge of systems architecture and with a dual perspective: defense and attack, both essential.

As noted in the comprehensive review \cite{mayoral2020robot}, robots are often shipped insecure, with defensive security mechanisms still in their early stages. The inherent complexity of robotic systems makes protection costly, and vendors frequently fail to take timely responsibility, extending zero-day exposure windows to several years on average. The urgency of this challenge has only intensified with recent advances in humanoid robotics. By 2020, researchers had already classified more than 1,000 vulnerabilities in robotic systems, yet the field remained fragmented, lacking both standardized assessment methodologies and comprehensive defensive frameworks. The introduction of specialized tools like the Robot Vulnerability Database (RVD) \cite{mayoral2022alurity} and offensive security frameworks \cite{mayoral2025offensive} represents significant progress, but the gap between theoretical models \cite{surve2024sok} and operational security remains vast. This contribution seeks to bridge this gap by providing empirical evidence of security mechanisms and vulnerabilities in a production humanoid platform.

\section{Motivation: The Humanoid Robot Revolution and Its Reality}

The year 2024 marked an inflection point in humanoid robotics, with unprecedented investment and media attention suggesting an imminent revolution. Tesla announced plans to produce 5,000 Optimus robots in 2025 \cite{investing2025tesla}, Figure secured BMW as a customer for its Figure 02 platform \cite{bmw2024figure}, and venture capital poured billions into the sector. Yet beneath this enthusiasm lies a more sobering reality: the market for humanoid robots remains almost entirely hypothetical, with even the most successful companies deploying only small numbers of robots in carefully controlled pilot projects.

Industry analysts and researchers suggest that meaningful deployment of humanoids in professional and public spaces may still be a decade away. Goldman Sachs projects the humanoid robot market will reach \$38 billion by 2035 \cite{goldmansachs2024humanoid}, while IDTechEx estimates \$30 billion in the same timeframe \cite{idtechex2025humanoid}---significant figures, but ones that acknowledge the lengthy development and adoption timeline ahead. As IEEE Spectrum noted in late 2024, future projections appear based on ``an extraordinarily broad interpretation of jobs that a capable, efficient, and safe humanoid robot---which does not currently exist---might conceivably be able to do'' \cite{spectrum2024humanoid}.

Despite this temporal gap between hype and reality, the proliferation of research and investment in advanced humanoid robots presents immediate cybersecurity challenges that demand attention. The Unitree G1, along with platforms from Tesla, Figure, Agility Robotics, and others detailed at \ref{app:companies}, represents a new generation of highly sophisticated machines combining biomimetic mechanical design with advanced artificial intelligence. These systems are being developed and tested today, establishing architectural patterns and security practices that will persist as the technology matures.

\subsection{Why Humanoid Robot Security Cannot Wait}

While widespread deployment may be years away, humanoid robots are already entering limited trials across critical sectors, establishing security precedents that will be difficult to change once deployed at scale:

\begin{itemize}
    \item \textbf{Industrial Applications}: Manufacturing, warehouse automation, and hazardous environment operations
    \item \textbf{Healthcare}: Patient care, rehabilitation assistance, and surgical support
    \item \textbf{Service Industries}: Customer service, hospitality, and retail assistance
    \item \textbf{Research and Education}: Human-robot interaction studies and STEM education platforms
    \item \textbf{Defense and Security}: Reconnaissance, search and rescue, and explosive ordnance disposal
\end{itemize}

The Unitree G1, a state-of-the-art humanoid platform, exemplifies the complexity of modern robotic systems that combine sophisticated hardware with intricate software architectures. With capabilities including:
\begin{itemize}
    \item 43 degrees of freedom for human-like movement
    \item Advanced computer vision with a RealSense camera
    \item Real-time motion planning and control
    \item Natural language processing and voice interaction
    \item Cloud connectivity for remote operation and updates advertised as ``Continuous OTA Software Upgrade and Update''
\end{itemize}

\subsection{The Unique Cybersecurity Challenges of Humanoids}

The convergence of several factors makes humanoid robot security a critical and distinct challenge:

\begin{enumerate}
    \item \textbf{Physical-Cyber Convergence}: Unlike traditional computing systems, compromised robots can cause physical harm. The safety-security nexus in robotics means that cybersecurity failures can directly translate to safety incidents \cite{kirschgens2018robot, mayoral2022alurity}.
    
    \item \textbf{Multi-modal Data Collection}: Humanoid robots integrate numerous sensors---cameras, microphones, LiDAR, force sensors---creating vast attack surfaces for data exfiltration and sensor spoofing attacks \cite{shin2017illusion, petit2015remote}.
    
    \item \textbf{Persistent Connectivity}: Modern humanoids maintain continuous connections to cloud services for telemetry, updates, and remote operation. Our analysis revealed persistent connections to external servers\footnote{Server addresses redacted for security purposes.}, transmitting detailed robot state information and in violation of the user's privacy.
    
    \item \textbf{Autonomous Decision-Making}: The integration of AI and machine learning creates new attack vectors through adversarial inputs and model manipulation. As highlighted in recent work on Cybersecurity AI \cite{mayoralvilches2025cybersecurityaidangerousgap}, the gap between automation and true autonomy introduces novel security considerations.
    
    \item \textbf{Supply Chain Complexity}: International manufacturing, third-party components, and diverse software stacks introduce multiple trust boundaries. The Unitree G1, for instance, combines Chinese hardware with open-source middleware (ROS 2, DDS) and proprietary control software.
\end{enumerate}

\section{Research Context and Contributions}

This work presents a comprehensive security assessment of the Unitree G1 humanoid robot, contributing to the nascent but critical field of humanoid robot cybersecurity. Building upon the foundations established by previous work in robot vulnerability assessment \cite{quarta2017experimental} and the emerging frameworks for offensive robot security \cite{mayoral2025offensive}, we provide empirical evidence of security mechanisms and vulnerabilities in a production humanoid platform.

Our investigation employs a multi-faceted approach combining:
\begin{itemize}
    \item Static analysis of the robot's filesystem and binaries
    \item Runtime observation of network communications and service interactions
    \item Cryptographic analysis of proprietary encryption mechanisms
    \item Systematic mapping of the service architecture and attack surface
\end{itemize}

This assessment reveals both sophisticated security measures---including a novel dual-layer encryption system and hierarchical service management---and critical vulnerabilities that have implications for the broader robotics industry. Most significantly, we demonstrate how theoretical security concerns translate into exploitable vulnerabilities in real-world humanoid systems.

\section{The Limits of Systematization in Emerging Fields}
The field of humanoid robot security faces a basic problem: how can we systematize knowledge that barely exists? Recent efforts to establish comprehensive security frameworks, such as the seven-layer model proposed by Surve et al.~\cite{surve2024sok}, provide valuable theoretical scaffolding. However, as the authors themselves acknowledge, ``no major, publicly documented cyberattacks have targeted humanoids to date.'' This absence of real-world incidents reveals a critical gap---we are attempting to categorize and defend against threats that remain largely hypothetical.

The reader must note that the systematization of knowledge (SoK) methodology, while valuable in mature fields with established attack patterns and defense mechanisms, encounters significant limitations when applied to nascent domains. In traditional cybersecurity contexts, SoK papers might synthesize decades of incidents, vulnerabilities, and countermeasures. For humanoid robotics, however, the empirical foundation is conspicuously absent. The 89 studies analyzed by Surve et al. predominantly focus on isolated vulnerabilities---LiDAR spoofing~\cite{shin2017illusion}, camera blinding~\cite{petit2015remote}, or acoustic gyroscope manipulation~\cite{son2015rocking}---rather than comprehensive system compromises observed in production environments.

This lack of real-world data creates a problem: security frameworks are being built without the actual evidence needed to prove their ideas work. The danger of premature systematization extends beyond academic exercises. As Quarta et al. demonstrated in their analysis of industrial robot controllers~\cite{quarta2017experimental}, theoretical vulnerabilities often pale in comparison to the implementation flaws discovered through hands-on investigation. Their empirical work revealed firmware backdoors and authentication bypasses that no amount of abstract modeling would have predicted.

The history of computer security demonstrates that robust defenses emerge from understanding actual attacks, not theoretical possibilities. The Morris Worm informed modern network security; Stuxnet reshaped industrial control system protection; WannaCry transformed patch management practices. For humanoid robotics to develop effective security measures, we need similar empirical foundations.

This is not to diminish the value of systematization efforts. Frameworks like the seven-layer model provide essential vocabulary and conceptual structure. However, they must be complemented---and ultimately validated---by deep technical investigations of actual systems. As previously discussed at \cite{mayoral2022alurity}, ``the gap between theoretical security models and operational robots is vast and growing.''

Our investigation of the Unitree G1 represents this empirical approach. Rather than theorizing about potential vulnerabilities, we:
\begin{itemize}
    \item Extracted and analyzed the robot, including its complete filesystem from production hardware
    \item Reverse engineered proprietary encryption implementations and system services
    \item Monitored runtime network communications and service interactions
    \item Documented actual data flows to external servers, in violation of the user's privacy
    \item Validated theoretical attack vectors through proof-of-concept implementations
    \item Documented the use of a dual-layer encryption system and hierarchical service management
\end{itemize}

This methodology yields actionable insights that abstract frameworks cannot provide. 

\section{Research Context and Contributions}

This work presents a comprehensive security assessment of the Unitree G1 humanoid robot, contributing to the nascent but critical field of humanoid robot cybersecurity. Building upon the foundations established by previous work in robot vulnerability assessment \cite{quarta2017experimental} and the emerging frameworks for offensive robot security \cite{mayoral2025offensive}, we provide empirical evidence of security mechanisms and vulnerabilities in a production humanoid platform. Our investigation employs a multi-faceted approach combining:
\begin{itemize}
    \item Physical teardown and inspection of the robot's hardware
    \item Static analysis of the robot's filesystem and binaries
    \item Runtime observation of network communications and service interactions
    \item Cryptographic analysis of proprietary encryption mechanisms
    \item Systematic mapping of the service architecture and attack surface
\end{itemize}

This assessment reveals both sophisticated security measures---including a novel dual-layer encryption system and hierarchical service management---and critical vulnerabilities that have implications for the broader robotics industry. Most significantly, we demonstrate how theoretical security concerns translate into exploitable vulnerabilities in real-world humanoid systems. Ultimately, we uncover how humanoids can be used as attack vectors in two ways: a) as trojan horses for data exfiltration and b) as a platform for system compromise.

\section{Document Organization}

This report is organized into five main chapters followed by detailed technical appendices:

\textbf{Main Chapters:}
\begin{itemize}
    \item \textbf{\autoref{chap:introduction}}: Establishes the motivation for humanoid robot security research and provides context for the investigation
    \item \textbf{\autoref{chap:architecture}}: Detailed analysis of the G1's system architecture, service orchestration, and communication infrastructure
    \item \textbf{\autoref{chap:cybersecurity}}: Comprehensive security assessment including methodology, encryption analysis, vulnerability findings, and recommendations
    \item \textbf{\autoref{chap:attack_vectors}}: Explores how humanoid robots can be exploited as trojan horses for data exfiltration and system compromise
    \item \textbf{\autoref{chap:conclusions}}: Summarizes findings, presents conclusions, and outlines future research directions
\end{itemize}

\textbf{Technical Appendices:}
\begin{itemize}
    \item \textbf{\autoref{chap:app_technical}}: Complete technical implementation details including reconstructed source code, configuration files, and cryptographic analysis tools
    \item \textbf{\autoref{chap:app_services}}: Comprehensive service architecture documentation with detailed enumeration of all system services and their interactions
    \item \textbf{\autoref{chap:app_tools}}: Collection of decryption tools, scripts, and utilities developed during the security assessment
    \item \textbf{\autoref{app:companies}}: Overview of the humanoid robotics industry landscape and major commercial platforms
\end{itemize}
\chapter{Humanoid Architecture}
\label{chap:architecture}

\section{Introduction}




This chapter provides a comprehensive reverse engineering and empirical analysis of the Unitree G1 humanoid robot's architecture.  By examining the actual implementation of service orchestration, inter-process communication, and external connectivity, we provide concrete evidence of how theoretical vulnerabilities translate into exploitable attack surfaces. Our analysis reveals previously undocumented persistent telemetry connections, proprietary encryption mechanisms, and architectural decisions that create cascading security implications across the robot's ecosystem. This empirical approach not only validates aspects of existing security models but also exposes implementation-specific vulnerabilities that theoretical frameworks cannot anticipate.

\section{Teardown of a Humanoid Robot}

To understand the physical attack surface and hardware vulnerabilities of modern humanoid robots, we performed a systematic teardown of the Unitree G1 platform. This analysis reveals the critical hardware components, their interconnections, and potential security implications arising from the physical implementation.

\subsection{Physical Disassembly Analysis}

The following figures document our progressive teardown of the Unitree G1 humanoid robot, revealing increasingly detailed views of the internal architecture and critical components.

\begin{figure}[h]
\centering
\includegraphics[width=0.8\textwidth]{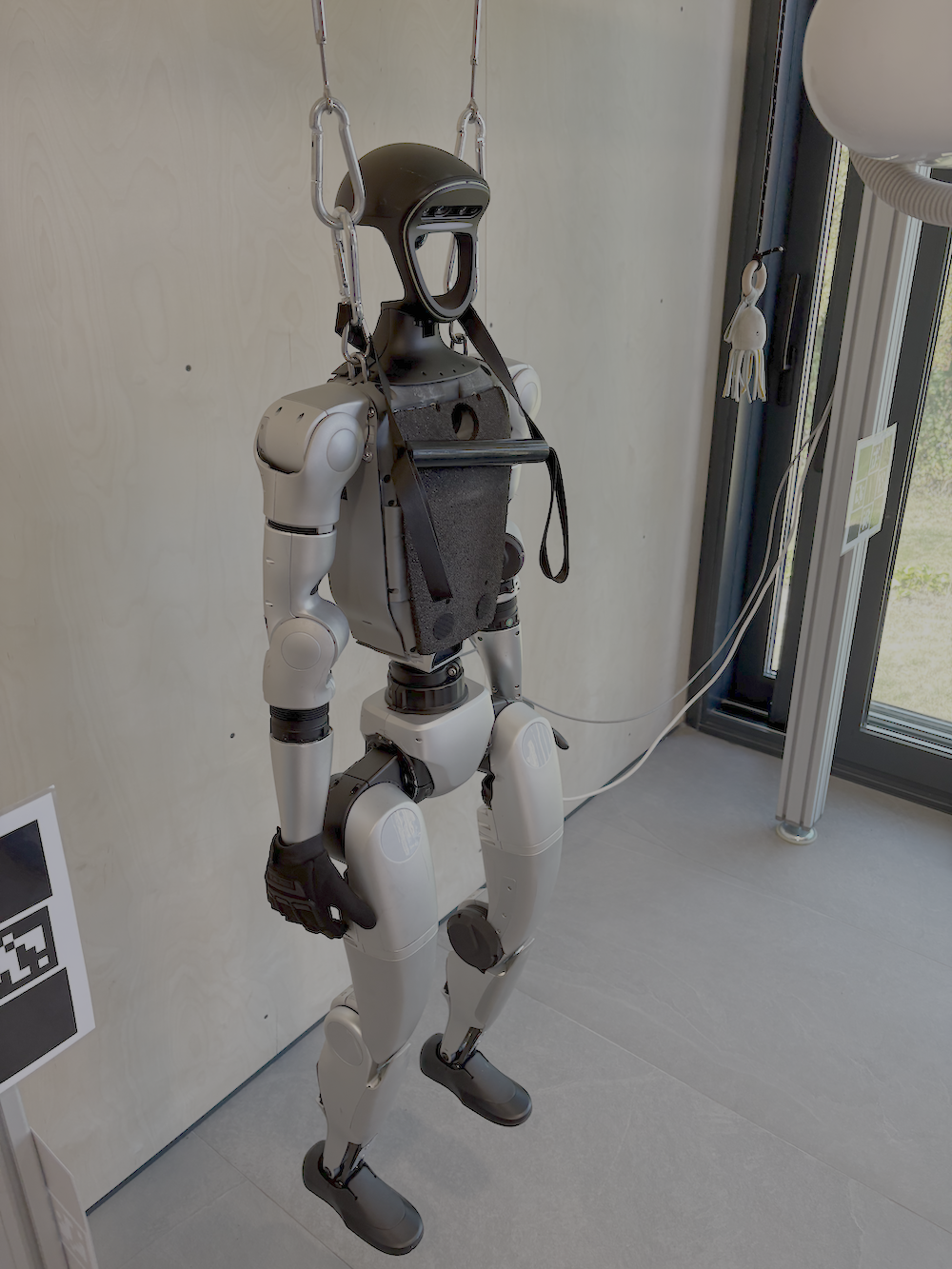}
\caption{\textbf{Full humanoid robot (Unitree G1) in support harness.} The external chassis shows no visible electronic components, presenting a sealed exterior designed to protect internal systems from environmental factors and physical tampering. The support harness mechanism is visible, used during testing and maintenance procedures.}
\label{fig:teardown1}
\end{figure}

\begin{figure}[h]
\centering
\includegraphics[width=0.8\textwidth]{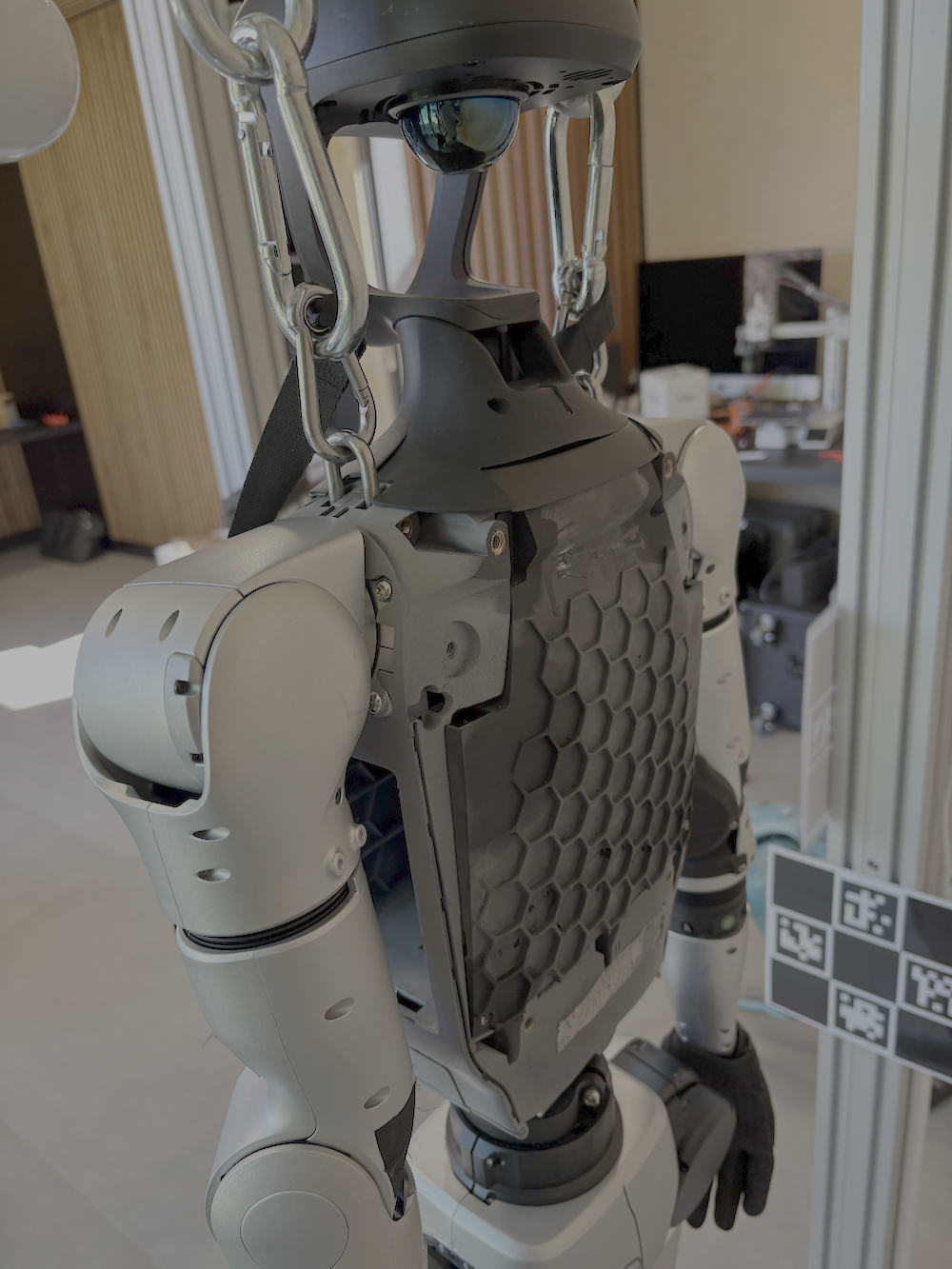}
\caption{\textbf{Upper torso with protective cover intact.} The chest plate serves as the primary access point to the robot's central compute and power management systems. External harness mounting points and sensor arrays are visible, but no internal electronics are exposed at this stage.}
\label{fig:teardown2}
\end{figure}

\begin{figure}[h]
\centering
\includegraphics[width=0.8\textwidth]{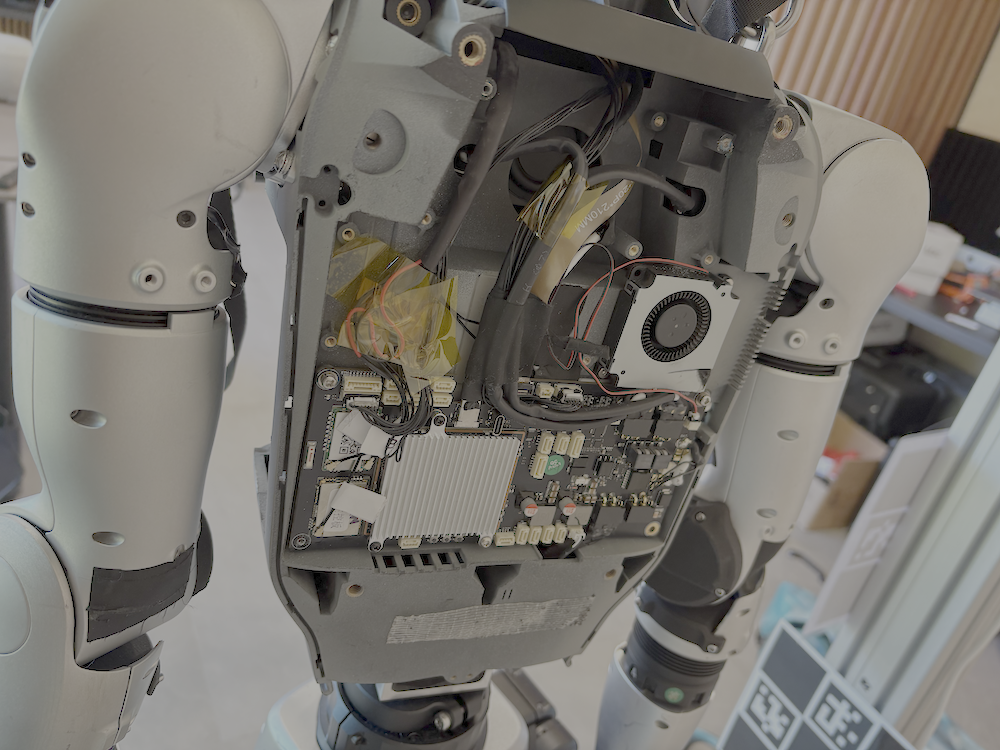}
\caption{\textbf{Chest cover opened revealing main PCB.} Visible components include: (1) Active cooling fan module for thermal management, (2) Main control board with multiple JST-style connectors, (3) Power distribution circuitry with visible capacitors, (4) Motor driver interfaces and sensor connection points. This PCB serves as the robot's central compute and power distribution hub, orchestrating all subsystem communications.}
\label{fig:teardown3}
\end{figure}

\begin{figure}[h]
\centering
\includegraphics[width=0.8\textwidth]{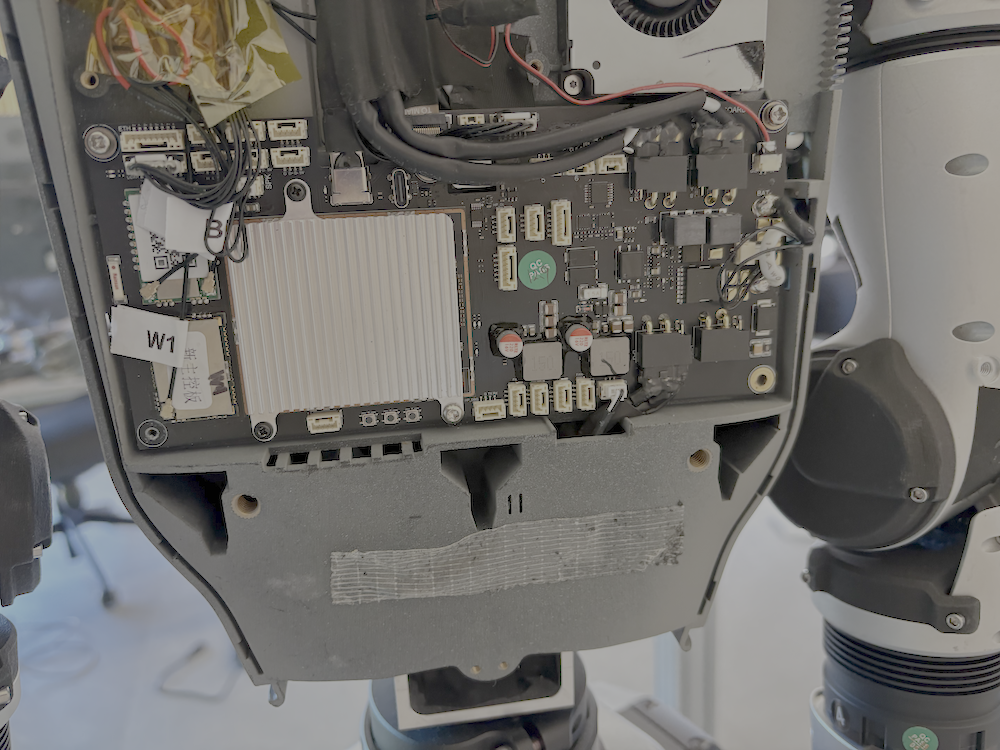}
\caption{\textbf{Close-up of main PCB architecture.} Key components visible: (1) Large heatsink covering the main SoC/CPU complex, (2) Multiple white JST connectors for modular connectivity, (3) Power regulation MOSFETs arranged in clusters, (4) Motor driver circuitry with dedicated power stages, (5) Sensor interface connectors for proprioceptive feedback. Internal wiring harness references suggest standardized connection protocols between subsystems.}
\label{fig:teardown4}
\end{figure}

\begin{figure}[h]
\centering
\includegraphics[width=0.8\textwidth]{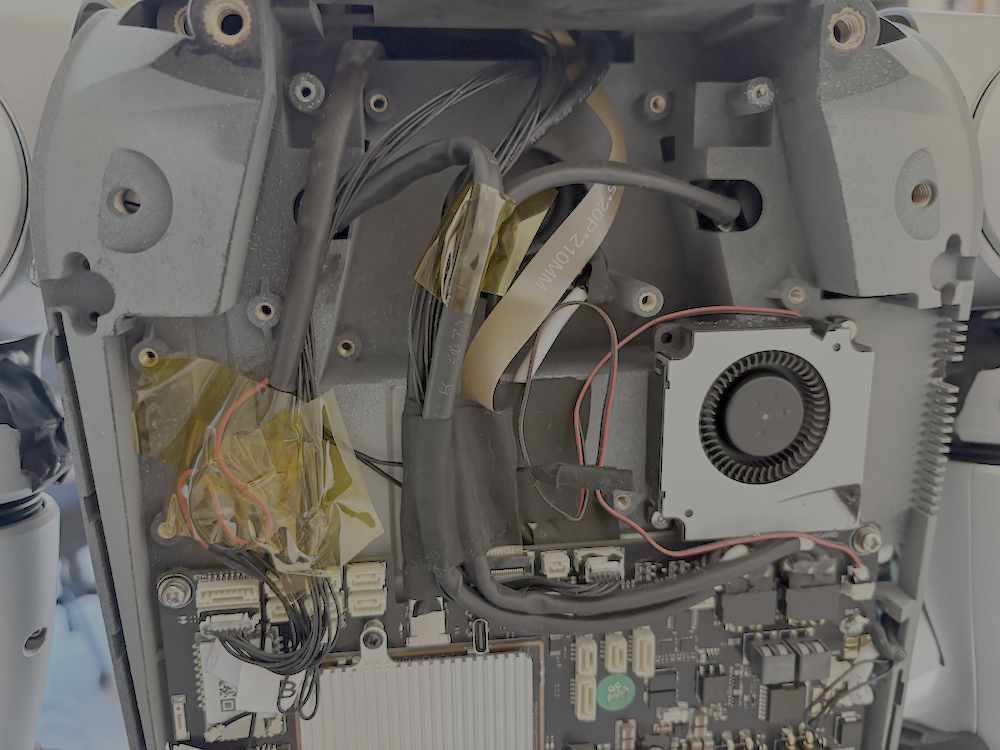}
\caption{\textbf{Upper PCB section focused on power management.} Critical power components: (1) High-current power MOSFETs for motor control, (2) Kapton tape insulation on high-voltage wire bundles, (3) Dedicated cooling fan for power section, (4) Multiple power regulation stages, (5) Bulk capacitors for power filtering and stability. This section handles the significant power requirements of humanoid locomotion and manipulation.}
\label{fig:teardown5}
\end{figure}

\begin{figure}[h]
\centering
\includegraphics[width=0.8\textwidth]{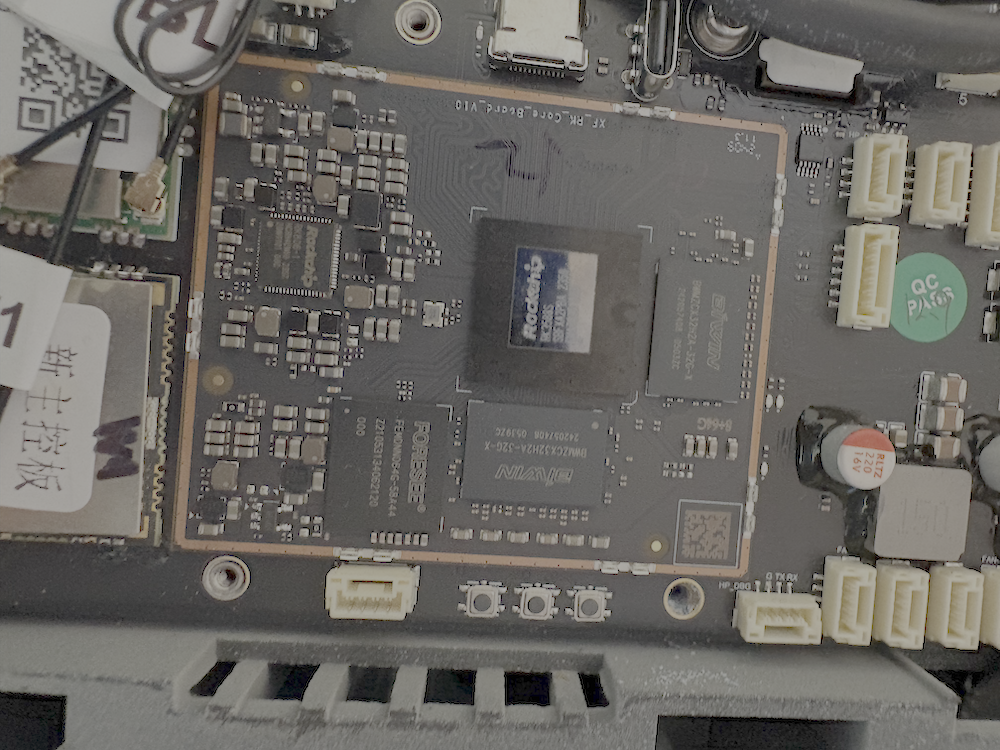}
\caption{\textbf{Macro view of main processing complex.} Silicon identification: (1) \textbf{Rockchip RK3588 SoC} - 8-core ARM Cortex-A76/A55 processor, (2) \textbf{FORESEE eMMC} - Embedded NAND flash storage, (3) \textbf{SK hynix LPDDR4/5 RAM} - High-bandwidth system memory, (4) \textbf{WiFi/BT Module} - Shielded RF section (left side), (5) \textbf{PMICs} - Multiple power management integrated circuits, (6) \textbf{270µF/16V capacitor} - Bulk power filtering. The RK3588 serves as the primary compute platform for high-level control, AI inference, and system orchestration.}
\label{fig:teardown6}
\end{figure}




\FloatBarrier

\subsection{Security Implications from Hardware Analysis}

The physical teardown reveals several security-relevant architectural decisions and potential vulnerability vectors:

\subsubsection{Attack Surface Identification}

The exposed hardware architecture presents multiple attack vectors:

\begin{enumerate}
\item \textbf{Physical Access Points}: The chest cavity provides direct access to the main control board, enabling hardware-level attacks if physical security is compromised.

\item \textbf{Modular Connectivity}: The extensive use of JST connectors, while facilitating maintenance, creates potential insertion points for malicious hardware implants or signal injection attacks.

\item \textbf{Unshielded Components}: Several critical ICs lack tamper-evident packaging or hardware security modules, potentially enabling chip-off attacks or side-channel analysis.

\item \textbf{Power Management Complexity}: The multi-stage power regulation system, while necessary for motor control, introduces potential fault injection opportunities through voltage glitching.
\end{enumerate}

\subsubsection{RK3588-Specific Vulnerabilities}

Our analysis of the Rockchip RK3588 SoC and its surrounding ecosystem revealed multiple security concerns documented in Table~\ref{tab:rk3588_vulns}.

\begin{landscape}
\begin{table}[h]
\centering
\caption{RK3588 Platform Security Vulnerabilities and Attack Vectors}
\label{tab:rk3588_vulns}
\footnotesize
\begin{tabular}{|p{2.5cm}|p{5cm}|p{4cm}|p{3cm}|p{4cm}|}
\hline
\textbf{ID / CVE} & \textbf{Description} & \textbf{Affects RK3588} & \textbf{Severity} & \textbf{Mitigation} \\ \hline

\textbf{CVE-2023-52660} & 
Improper error handling in \texttt{rkisp1} driver ISP routines allows local DoS through malformed media device operations & 
Yes: Core driver for RK3588 camera/video subsystem & 
Medium: Local DoS potential & 
Patch kernel driver; restrict media device access to trusted processes \\ \hline

\textbf{CVE-2025-38081} & 
Out-of-bounds register access in \texttt{spi-rockchip} driver when using GPIO chip selects with indices beyond hardware limits & 
Yes: Affects all RK3588 SPI interfaces & 
High: Memory corruption, potential privilege escalation & 
Apply kernel patches; validate GPIO CS configuration; update to patched kernel version \\ \hline

\textbf{Secure Boot Weaknesses} & 
Limited public documentation on RK3588 secure boot implementation; reverse-engineered ``ramboot'' component shows exploitable gaps in chain of trust & 
Critical: Boot security is fundamental to platform integrity & 
Critical: Complete firmware compromise possible & 
Enable full secure boot chain; protect signing keys; audit bootloader implementation; monitor vendor security advisories \\ \hline

\textbf{TEE Documentation Gaps} & 
TrustedFirmware-A supports RK3588 but threat model parameters incomplete; security boundaries poorly documented & 
Yes: TEE is present but configuration uncertain & 
Medium: Misconfiguration risks & 
Use latest TF-A version; conduct security audit; implement defense-in-depth beyond TEE \\ \hline

\textbf{CVE-2024-57256} & 
U-Boot filesystem vulnerabilities allow malicious storage modifications to execute untrusted code before verified boot engages & 
Yes if using vulnerable U-Boot versions & 
Critical: Pre-boot compromise & 
Enable verified boot; cryptographically protect filesystem; update U-Boot; secure physical storage access \\ \hline

\end{tabular}
\end{table}
\end{landscape}

\subsubsection{Implications for Humanoid Security}

The hardware teardown and vulnerability analysis reveal critical security considerations for humanoid robot deployments:

\begin{securityfinding}{Hardware-Based Attack Vectors}
The combination of accessible physical interfaces, documented SoC vulnerabilities, and limited secure boot implementations creates a compound risk profile. An attacker with physical access could potentially:
\begin{itemize}
\item Extract firmware and cryptographic materials through unprotected debug interfaces
\item Inject malicious code at the bootloader level, bypassing higher-level security controls
\item Perform side-channel attacks on the unshielded RK3588 during cryptographic operations
\item Compromise the robot's trusted execution environment through documented TEE weaknesses
\end{itemize}
\end{securityfinding}

These hardware-level vulnerabilities serve as foundational attack vectors that can undermine the entire security architecture described in subsequent sections. The lack of hardware security modules, combined with the RK3588's known vulnerabilities, suggests that physical security must be a primary consideration in any deployment scenario.

The transition from hardware to software security boundaries occurs at multiple levels---bootloader, kernel, and userspace---each inheriting the security weaknesses of the layer below. This cascading vulnerability model, empirically validated through our teardown, confirms theoretical predictions about cross-layer attack propagation in humanoid systems while revealing implementation-specific weaknesses that purely theoretical analyses would miss.

\section{High-Level Systems Architecture}
After an initial hardware analysis, we turn our attention to the software architecture of the Unitree G1 humanoid robot. Based on our observations, Figure~\ref{fig:hl_arch} summarizes the ecosystem architecture. The system comprises three primary domains: Cloud Services (including MQTT server, STUN/TURN service, and HTTP Web API), the mobile app/external interfaces (with WebRTC and BLE modules), and the robot itself with dual PC architecture. Communication paths include WebRTC data channels (with signaling via STUN/TURN and HTTP), BLE connections to \texttt{upper\_bluetooth}, DDS/RTPS on base ports 7400/7401, and MQTT publish/subscribe channels. The master service orchestrates this ecosystem through layered configuration protection and service inventory management.

\begin{landscape}
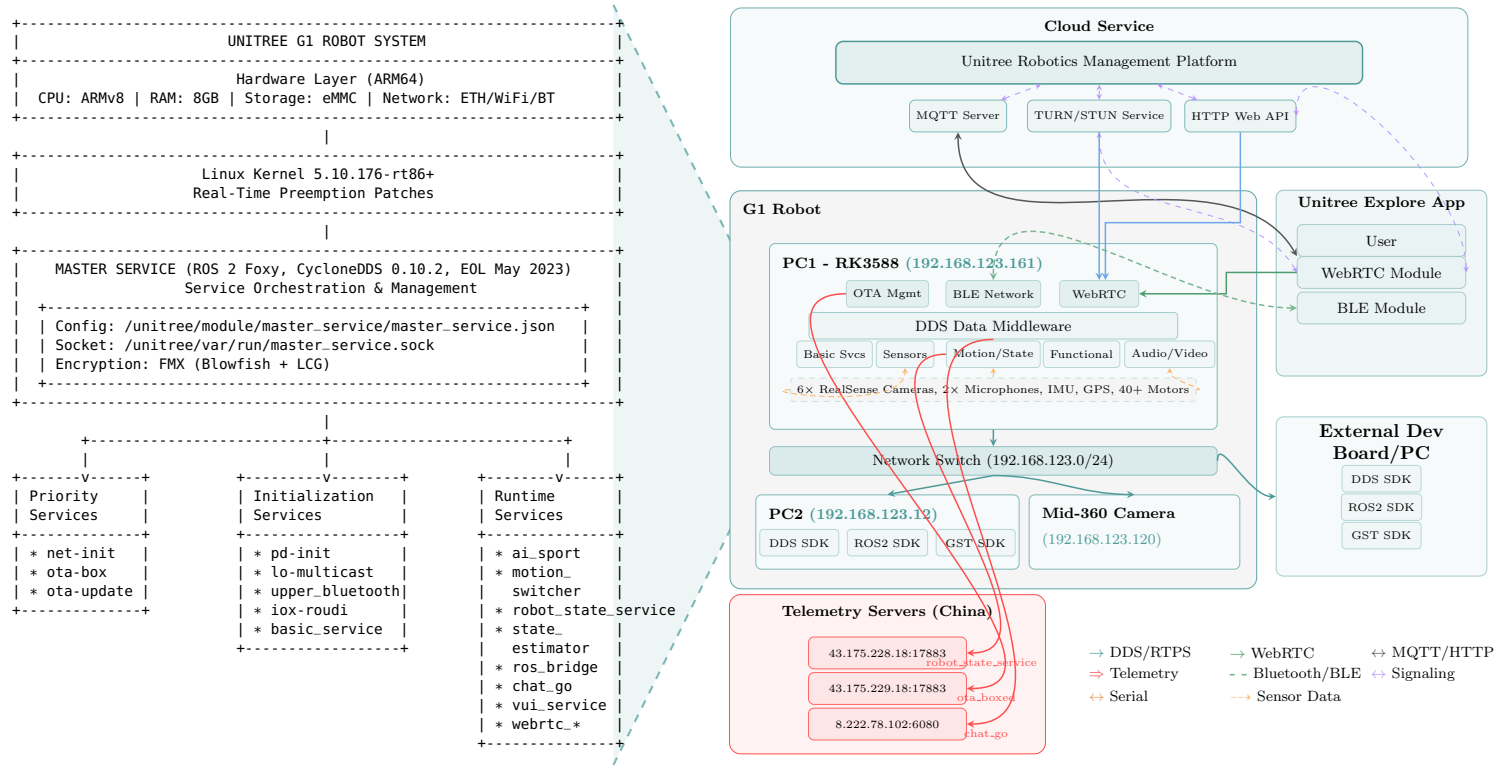
\begin{figure}[h]
    \centering
    \begin{tikzpicture}[overlay, remember picture]
        \coordinate (left-start) at (-2.5cm, 0cm);
        \coordinate (right-end) at (0cm, -5cm);
        \coordinate (left-start-2) at (-2.5cm, -10cm);
        \coordinate (right-end-2) at (0cm, -5cm);
        
        \fill[unitree_primary, opacity=0.1] 
            (left-start) -- (right-end) -- (right-end-2) -- (left-start-2) -- cycle;
        
        \draw[dashed, thick, unitree_primary, opacity=0.7] 
            (left-start) -- (right-end);
        \draw[dashed, thick, unitree_primary, opacity=0.7] 
            (left-start-2) -- (right-end-2);
    \end{tikzpicture}
    
    \begin{subfigure}[c]{0.3\textwidth}
        \centering
        \begin{Verbatim}[fontsize=\tiny]
+---------------------------------------------------------------------+
|                        UNITREE G1 ROBOT SYSTEM                      |
+---------------------------------------------------------------------+
|                         Hardware Layer (ARM64)                      |
|  CPU: ARMv8 | RAM: 8GB | Storage: eMMC | Network: ETH/WiFi/BT       |
+---------------------------------------------------------------------+
                                    |
+---------------------------------------------------------------------+
|                     Linux Kernel 5.10.176-rt86+                     |
|                    Real-Time Preemption Patches                     |
+---------------------------------------------------------------------+
                                    |
+---------------------------------------------------------------------+
|    MASTER SERVICE (ROS 2 Foxy, CycloneDDS 0.10.2, EOL May 2023)     |
|                   Service Orchestration & Management                |
|  +--------------------------------------------------------------+   |
|  | Config: /unitree/module/master_service/master_service.json   |   |
|  | Socket: /unitree/var/run/master_service.sock                 |   |
|  | Encryption: FMX (Blowfish + LCG)                             |   |
|  +--------------------------------------------------------------+   |
+---------------------------------------------------------------------+
                                    |
        +---------------------------+---------------------------+
        |                           |                           |
+-------v------+          +---------v--------+        +--------v------+
| Priority     |          | Initialization   |        | Runtime       |
| Services     |          | Services         |        | Services      |
+--------------+          +------------------+        +---------------+
| * net-init   |          | * pd-init        |        | * ai_sport    |
| * ota-box    |          | * lo-multicast   |        | * motion_     |
| * ota-update |          | * upper_bluetooth|        |   switcher    |
+--------------+          | * iox-roudi      |        | * robot_state_service
                          | * basic_service  |        | * state_      |
                          +------------------+        |   estimator   |
                                                      | * ros_bridge  |
                                                      | * chat_go     |
                                                      | * vui_service |
                                                      | * webrtc_*    |
                                                      +---------------+
        \end{Verbatim}
        \label{fig:hl_arch_internal}
    \end{subfigure}
    \hfill
    \begin{subfigure}[c]{0.95\textwidth}
        \centering
        \includegraphics[width=\textwidth]{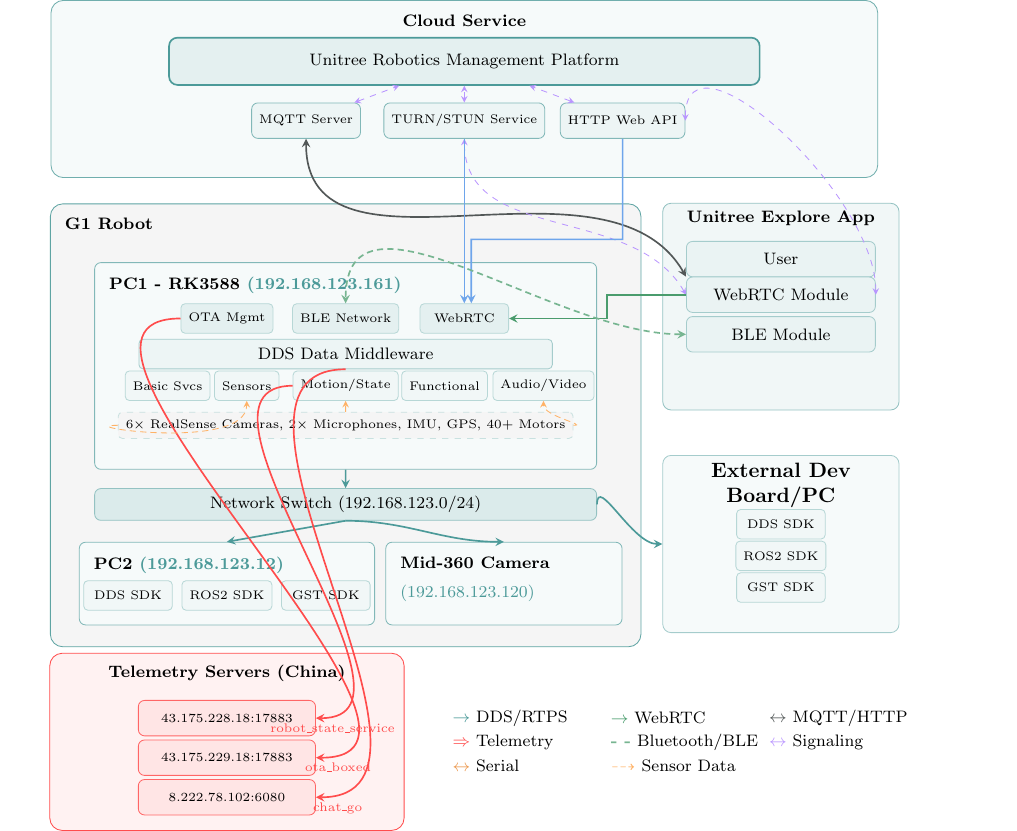}
        \label{fig:hl_arch_external}
    \end{subfigure}
    \caption{Humanoid Robot Architecture: Internal system structure showing hardware layer, Linux kernel, master service orchestration, and service hierarchy (left), and high-level ecosystem with communication paths showing authorized cloud services, telemetry servers, and internal components including obstacle avoidance, path planning, and speech recognition with DDS/ROS2 compatibility (right). \textcolor{red!70}{Critical finding: Persistent telemetry connections to external servers transmit robot state and sensor data without explicit user consent.}}
    \label{fig:hl_arch}
\end{figure}
\end{landscape}    

\section{Internal Service Architecture}

\subsubsection{System Overview}

The following ASCII diagram shows the high-level Unitree G1 system layout captured during runtime introspection (Ubuntu 20.04.5 LTS; Linux 5.10.176-rt86+, ARM64). It highlights the hardware base, soft real-time kernel, and the master service that orchestrates Priority, Initialization, and Runtime services based on our system architecture analysis and runtime observations.

\begin{Verbatim}[fontsize=\small]
+---------------------------------------------------------------------+
|                        UNITREE G1 ROBOT SYSTEM                      |
+---------------------------------------------------------------------+
|                         Hardware Layer (ARM64)                      |
|  CPU: ARMv8 | RAM: 8GB | Storage: eMMC | Network: ETH/WiFi/BT       |
+---------------------------------------------------------------------+
                                    |
+---------------------------------------------------------------------+
|                     Linux Kernel 5.10.176-rt86+                     |
|                    Real-Time Preemption Patches                     |
+---------------------------------------------------------------------+
                                    |
+---------------------------------------------------------------------+
|   MASTER SERVICE (ROS 2 Foxy, CycloneDDS 0.10.2, EOL May 2023)      |
|                   Service Orchestration & Management                |
|  +--------------------------------------------------------------+   |
|  | Config: /unitree/module/master_service/master_service.json   |   |
|  | Socket: /unitree/var/run/master_service.sock                 |   |
|  | Encryption: FMX (Blowfish + LCG)                             |   |
|  +--------------------------------------------------------------+   |
+---------------------------------------------------------------------+
                                    |
        +---------------------------+---------------------------+
        |                           |                           |
+-------v------+          +---------v--------+        +--------v------+
| Priority     |          | Initialization   |        | Runtime       |
| Services     |          | Services         |        | Services      |
+--------------+          +------------------+        +---------------+
| * net-init   |          | * pd-init        |        | * ai_sport    |
| * ota-box    |          | * lo-multicast   |        | * motion_     |
| * ota-update |          | * upper_bluetooth|        |   switcher    |
+--------------+          | * iox-roudi      |        | * robot_state_service
                          | * basic_service  |        | * state_      |
                          +------------------+        |   estimator   |
                                                      | * ros_bridge  |
                                                      | * chat_go     |
                                                      | * vui_service |
                                                      | * webrtc_*    |
                                                      +---------------+
\end{Verbatim}

\subsubsection{Communication Middleware Stack}

The G1 uses a multi-protocol middleware: DDS/Iceoryx for high-throughput IPC, a ROS~2 Foxy powered by CycloneDDS (released June 2020, EOL May 2023 - outdated and unsupported), and a WebRTC stack (signal server on port~8081). Shared memory transport is visible under \texttt{/dev/shm/iceoryx\_*}. Our network analysis confirmed active endpoints and communication patterns.

Something worth noting is that the G1 uses a ROS 2 Foxy (released June 2020, EOL May 2023 - outdated and unsupported) powered by CycloneDDS 0.10.2, which dates from approximately 2022, missing 3+ version releases.

\begin{Verbatim}[fontsize=\small]
+---------------------------------------------------------------------+
|                     Communication Infrastructure                    |
+---------------------------------------------------------------------+
|                                                                     |
|  +----------------+  +----------------+  +----------------+         |
|  |   DDS/Iceoryx  |  |    ROS 2 Foxy  |  |    WebRTC      |         |
|  |   (iox-roudi)  |  |  (CycloneDDS   |  |   Bridge       |         |
|  |   Port: 7400   |  |      0.10.2)   |  |  Port: 8081    |         |
|  +----------------+  +----------------+  +----------------+         |
|         |                    |                    |                 |
|         +--------------------+--------------------+                 |
|                              |                                      |
|                    Shared Memory IPC                                |
|                 (/dev/shm/iceoryx_*)                                |
+---------------------------------------------------------------------+
\end{Verbatim}

\subsubsection{Core Services Architecture}

The motion stack centers on \texttt{ai\_sport} (primary controller), supported by \texttt{state\_estimator}, \texttt{motion\_switcher}, and the \texttt{robot\_state} broadcaster. Arms are managed by \texttt{dex3\_service\_l/r}. Resource usage measurements confirm these values from our telemetry analysis.

\paragraph{1. Motion Control Subsystem}
\begin{Verbatim}[fontsize=\small]
+---------------------------------------------------------------------+
|                        Motion Control Stack                         |
+---------------------------------------------------------------------+
|                                                                     |
|  +-----------------------------------------+                        |
|  |        ai_sport (PID 1603)              |                        |
|  |   CPU: 145% | Mem: 135MB                |                        |
|  |   Primary motion planning & control     |                        |
|  +-----------------------------------------+                        |
|                      |                                              |
|  +---------------------+---------------------+                      |
|  |                     |                      |                     |
|  v                     v                      v                     |
| motion_switcher   state_estimator      robot_state                  |
|  (PID 1164)        (PID 1304)          (PID 1225)                   |
|  CPU: 3.2%         CPU: 30.4%          CPU: 5.1%                    |
|                                                                     |
|  +----------------+  +----------------+                             |
|  | dex3_service_l |  | dex3_service_r |  (Arm controllers)          |
|  +----------------+  +----------------+                             |
|                                                                     |
|  +----------------+                                                 |
|  | g1_arm_example |  (Example arm control service)                  |
|  +----------------+                                                 |
+---------------------------------------------------------------------+
\end{Verbatim}

\paragraph{2. Human–Machine Interface}
Voice and chat services operate continuously; memory usage of \texttt{vui\_service} aligns with microphone capture observed during our telemetry assessment. Conversational back-end traffic for \texttt{chat\_go} is visible in the snapshot collected (port~6080 to 8.222.78.102).

\begin{Verbatim}[fontsize=\small]
+---------------------------------------------------------------------+
|                    Human-Machine Interface (HMI)                    |
+---------------------------------------------------------------------+
|                                                                     |
|  +------------------------------------------+                       |
|  |         vui_service (PID 1413)           |                       |
|  |    Voice User Interface (14.2% Memory)   |                       |
|  |    /unitree/module/vui_service/          |                       |
|  +------------------------------------------+                       |
|                                                                     |
|  +------------------------------------------+                       |
|  |          chat_go (PID 1069)              |                       |
|  |     Natural Language Processing          |                       |
|  |     Python-based service                 |                       |
|  +------------------------------------------+                       |
|                                                                     |
|  +------------------------------------------+                       |
|  |         video_hub (disabled)             |                       |
|  |     Video streaming service              |                       |
|  +------------------------------------------+                       |
+---------------------------------------------------------------------+
\end{Verbatim}

\paragraph{3. Connectivity Services}
The WebRTC bridge cluster (master, multicast responder, and signal server) exposes local signaling on port~8081. Bluetooth stack combines \texttt{bluetoothd}, \texttt{btgatt\_server}, and an \texttt{upper\_bluetooth} helper.

\begin{Verbatim}[fontsize=\small]
+---------------------------------------------------------------------+
|                        Connectivity Services                        |
+---------------------------------------------------------------------+
|                                                                     |
|  +------------------------------------------+                       |
|  |        WebRTC Bridge (PID 1431)          |                       |
|  |   webrtc_bridge (Master)                 |                       |
|  |   webrtc_multicast_responder (PID 1449)  |                       |
|  |   webrtc_signal_server (PID 1465)        |                       |
|  |   Port: 8081 (Signal Server)             |                       |
|  +------------------------------------------+                       |
|                                                                     |
|  +------------------------------------------+                       |
|  |      Bluetooth Services                  |                       |
|  |   bluetoothd (PID 466)                   |                       |
|  |   btgatt-server (PID 765, 913)           |                       |
|  |   upper_bluetooth service                |                       |
|  +------------------------------------------+                       |
|                                                                     |
|  +------------------------------------------+                       |
|  |      Network Management                  |                       |
|  |   net_switcher service                   |                       |
|  |   Interfaces: eth0, wlan0                |                       |
|  |   IPs: 192.168.123.161, 192.168.8.193    |                       |
|  +------------------------------------------+                       |
+---------------------------------------------------------------------+
\end{Verbatim}

\paragraph{4. System Services}
OTA components (\texttt{ota\_box}, \texttt{ota\_update}) and \texttt{basic\_service} are supervised by the master service. Live TCP sessions show persistent links to external servers on port 17883 (\texttt{ota\_boxed}, \texttt{robot\_state\_service}).

\begin{Verbatim}[fontsize=\small]
+---------------------------------------------------------------------+
|                         System Services                             |
+---------------------------------------------------------------------+
|                                                                     |
|  +------------------------------------------+                       |
|  |      OTA Update System                   |                       |
|  |   ota_box service                        |                       |
|  |   ota_update service                     |                       |
|  |   Socket: /unitree/var/run/ota_boxed.sock|                       |
|  +------------------------------------------+                       |
|                                                                     |
|  +------------------------------------------+                       |
|  |      Basic Service                       |                       |
|  |   System initialization                  |                       |
|  |   Hardware management                    |                       |
|  +------------------------------------------+                       |
|                                                                     |
|  +------------------------------------------+                       |
|  |      Bashrunner Service                  |                       |
|  |   Script execution framework              |                      |
|  +------------------------------------------+                       |
+---------------------------------------------------------------------+
\end{Verbatim}

\subsubsection{Hardware Interfaces}

Observed devices include six video nodes, multiple I2C buses, and Ethernet/Wi-Fi NICs. Robot identifiers (codes, MACs) and hardware constants are documented in the evidence appendix.

\begin{Verbatim}[fontsize=\small]
+---------------------------------------------------------------------+
|                        Hardware Interfaces                          |
+---------------------------------------------------------------------+
|                                                                     |
|  Cameras:                                                           |
|  * /dev/video0-5 (6 video devices)                                  |
|                                                                     |
|  I2C Buses:                                                         |
|  * /dev/i2c-0, i2c-2, i2c-4, i2c-6                                  |
|                                                                     |
|  Network:                                                           |
|  * eth0: 192.168.123.161/24 (MAC: 7e:1d:75:60:f5:89)                |
|  * wlan0: 192.168.8.193/24 (MAC: 78:22:88:a7:41:ed)                 |
|                                                                     |
+---------------------------------------------------------------------+
\end{Verbatim}

\subsubsection{Service Launch Sequence}

Launch ordering as instantiated by \texttt{master\_service}: Priority → Initialization → Runtime. This aligns with process trees and logs from our security assessment.

\begin{Verbatim}[fontsize=\small]
Start
  |
  +-> master_service
  |     |
  |     +-> Priority Services (net-init, ota-box, ota-update)
  |     |
  |     +-> Initialization Services
  |     |     +-> pd-init
  |     |     +-> lo-multicast
  |     |     +-> upper_bluetooth
  |     |     +-> iox-roudi (DDS)
  |     |     +-> basic_service
  |     |
  |     +-> Runtime Services
  |           +-> motion_switcher
  |           +-> state_estimator
  |           +-> robot_state
  |           +-> ai_sport
  |           +-> ros_bridge
  |           +-> dex3_service_l/r
  |           +-> g1_arm_example
  |           +-> chat_go
  |           +-> vui_service
  |           +-> webrtc_bridge
  |           +-> net_switcher
  |           +-> bashrunner
  |
  +-> System Ready
\end{Verbatim}

\subsubsection{Inter-Process Communication}

The platform relies on multiple mechanisms.

\paragraph{Unix domain sockets}
\begin{itemize}
  \item \texttt{/unitree/var/run/master\_service.sock} — master control RPC
  \item \texttt{/unitree/var/run/ota\_boxed.sock} — OTA management
\end{itemize}

\paragraph{DDS/Iceoryx shared memory}
\begin{itemize}
  \item \texttt{/dev/shm/iceoryx\_*} — high-performance data sharing
  \item UDP port~7400 — discovery traffic
\end{itemize}

\paragraph{ROS~2}
\begin{itemize}
  \item ROS 2 Foxy, CycloneDDS 0.10.2, EOL May 2023 - outdated and unsupported
  \item DDS version 0.10.2 which is missing 3+ releases
\end{itemize}

\paragraph{WebRTC signaling}
\begin{itemize}
  \item Port~8081 signal server; multicast responder for discovery
\end{itemize}

\subsubsection{Security Architecture}

Dual-layer configuration protection (``FMX'') combines a Blowfish layer and a device-bound LCG transform. Process hardening includes self-\texttt{ptrace} and supervised restarts.

\begin{Verbatim}[fontsize=\small]
+---------------------------------------------------------------------+
|                       Security Architecture                         |
+---------------------------------------------------------------------+
|                                                                     |
|  Configuration Protection:                                          |
|  * FMX Encryption (Dual-layer)                                      |
|    - Layer 2: Blowfish ECB (128-bit key)                            |
|    - Layer 1: LCG Stream Cipher (w/ hardware-bound seed)            |
|                                                                     |
|  Process Protection:                                                |
|  * Self-ptrace anti-debugging                                       |
|  * Service monitoring & auto-restart                                |
|  * Maximum 30 protection restarts                                   |
|                                                                     |
|  Network Security:                                                  |
|  * SSH (Port 22) - Disabled by service                              |
|  * WebRTC encryption for remote access                              |
|  * Bluetooth GATT security                                          |
+---------------------------------------------------------------------+
\end{Verbatim}

\subsubsection{Performance Metrics}

Representative CPU and memory usage of key services (aggregated from runtime snapshots):

\paragraph{CPU (top consumers)}
\begin{itemize}
  \item \texttt{ai\_sport}: 145\% (multi-core)
  \item \texttt{state\_estimator}: 30.4\%
  \item \texttt{vui\_service}: 6.1\%
  \item \texttt{robot\_state*}: 5.1\%
  \item \texttt{webrtc\_bridge}: 5.2\%
\end{itemize}

\paragraph{Memory (top consumers)}
\begin{itemize}
  \item \texttt{vui\_service}: 1.15\,GB (14.2\%)
  \item \texttt{ai\_sport}: 136\,MB (1.6\%)
  \item \texttt{chat\_go}: 70\,MB (0.8\%)
  \item \texttt{ros\_bridge}: 20\,MB (0.2\%)
\end{itemize}

\subsubsection{File System Layout}

Key directories for binaries, configuration, runtime sockets, and identifiers:

\begin{Verbatim}[fontsize=\small]
/unitree/
+-- module/          # Service binaries
|   +-- master_service/
|   +-- ai_sport/
|   +-- motion_switcher/
|   +-- robot_state/
|   +-- state_estimator/
|   +-- ...
+-- etc/            # Configuration
|   +-- master_service/
|       +-- service/
|       +-- cmd/
+-- var/
|   +-- run/        # Runtime sockets
+-- sbin/           # System binaries
|   +-- iox-roudi
|   +-- mscli
+-- robot/
    +-- basic/      # Hardware identifiers
\end{Verbatim}

\begin{mdframed}[backgroundcolor=unitree_light!40,roundcorner=6pt]
\textbf{Live telemetry.} Snapshot shows persistent TCP sessions to external servers on port 17883 by \texttt{robot\_state\_service} and \texttt{ota\_boxed}, and a \texttt{chat\_go} connection on port 6080.
\end{mdframed}

\chapter{Cybersecurity of Humanoids}
\label{chap:cybersecurity}

\begin{mdframed}[backgroundcolor=unitree_light!40,roundcorner=6pt]

Details regarding initial access methods to the Unitree G1 robot system are intentionally withheld from this report for safety and security reasons. The techniques used to gain initial system access involve sensitive information that could potentially be misused if publicly disclosed. Our analysis focuses on the security architecture and vulnerabilities discovered after obtaining research access in our own robot.
\end{mdframed}

\section{System Architecture Analysis}
\label{sec:architecture}

\subsection{Process Hierarchy Discovery}

\begin{figure}[h]
\centering
\begin{tikzpicture}[scale=0.7, transform shape,
    master/.style={rectangle, draw=unitree_primary, fill=unitree_primary!20, text width=2.8cm, align=center, minimum height=0.8cm, rounded corners, thick, font=\small\bfseries},
    category/.style={rectangle, draw=unitree_accent, fill=unitree_accent!10, text width=2.2cm, align=center, minimum height=0.6cm, rounded corners, font=\small\bfseries},
    service/.style={rectangle, draw=unitree_secondary, fill=unitree_secondary!5, text width=1.8cm, align=center, minimum height=0.5cm, font=\scriptsize, rounded corners=2pt},
    comm/.style={rectangle, draw=unitree_orange, fill=unitree_orange!5, text width=1.8cm, align=center, minimum height=0.5cm, font=\scriptsize, rounded corners=2pt}
]
    \node[master] (master) at (0,0) {master\_service\\{\small (PID managed)}};
    
    \node[category] (priority) at (-6, -2.5) {Priority\\Services};
    \node[category] (runtime) at (0, -2.5) {Runtime\\Services};
    \node[category] (optional) at (6, -2.5) {Optional\\Services};
    
    \node[service] at (-7.2, -4.2) {iox-roudi\\{\tiny IPC daemon}};
    \node[service] at (-4.8, -4.2) {basic\_service\\{\tiny HW control}};
    
    \node[service] at (-2.4, -4.2) {ai\_sport\\{\tiny Motion AI}};
    \node[service] at (0, -4.2) {state\_estimator\\{\tiny Sensor fusion}};
    \node[service] at (2.4, -4.2) {robot\_state\\{\tiny State mgmt}};
    \node[service] at (-1.2, -5.8) {motion\_switcher\\{\tiny Mode control}};
    \node[service] at (1.2, -5.8) {ros\_bridge\\{\tiny ROS2 bridge}};
    
    \node[service] at (4.8, -4.2) {g1\_arm\_example\\{\tiny Arm demo}};
    \node[service] at (7.2, -4.2) {auto\_test\_arm\\{\tiny Arm test}};
    \node[service] at (6, -5.8) {auto\_test\_low\\{\tiny Low test}};
    
    \node[category] (comm_cat) at (0, -7.5) {Communication\\Services};
    
    \node[comm] at (-4.8, -9.2) {chat\_go\\{\tiny Voice AI}};
    \node[comm] at (-2.4, -9.2) {vui\_service\\{\tiny Voice UI}};
    \node[comm] at (0, -9.2) {webrtc\_*\\{\tiny WebRTC}};
    \node[comm] at (2.4, -9.2) {video\_hub\\{\tiny Video stream}};
    \node[comm] at (4.8, -9.2) {net\_switcher\\{\tiny Network}};
    
    \node[comm] at (-2.4, -10.8) {upper\_bluetooth\\{\tiny BT service}};
    \node[comm] at (0, -10.8) {bashrunner\\{\tiny Script exec}};
    \node[comm] at (2.4, -10.8) {ota\_box\\{\tiny Updates}};
    
    \draw[thick, unitree_primary, -] (master.south) -- ++(0,-0.8) coordinate (branch);
    \draw[thick, unitree_primary] (branch) -- ++(-6,0) -- (priority.north);
    \draw[thick, unitree_primary] (branch) -- (runtime.north);
    \draw[thick, unitree_primary] (branch) -- ++(6,0) -- (optional.north);
    
    \draw[thick, unitree_accent, dashed] (0, -6.5) -- (comm_cat.north);
    
    \node[font=\tiny, text=gray] at (8, -11) {22 services total};
\end{tikzpicture}
\caption{Process hierarchy of the Unitree G1 robot showing service management architecture}
\label{fig:process_hierarchy}
\end{figure}
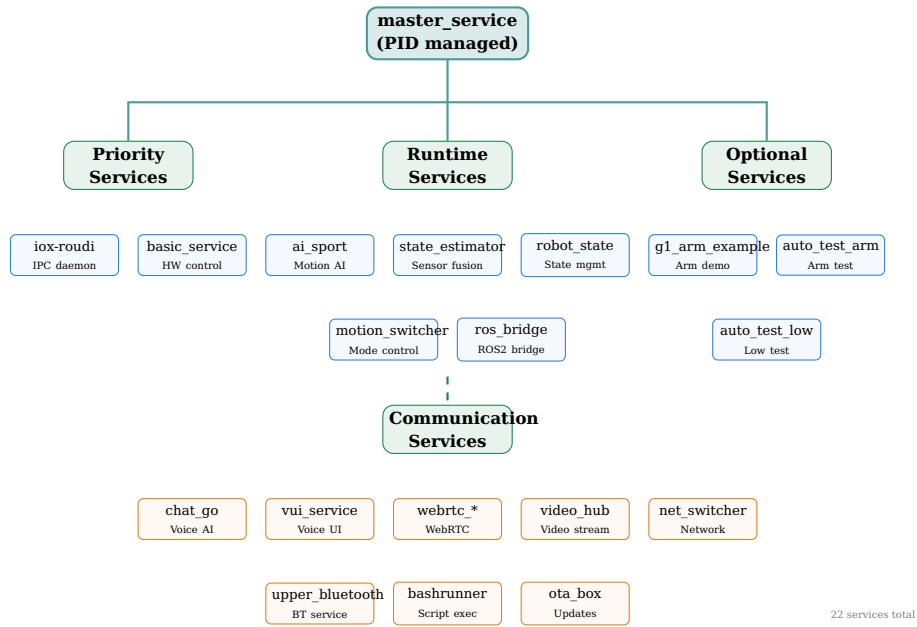

Initial investigation revealed a sophisticated service management architecture centered around the \texttt{master\_service} binary and as described in \autoref{fig:process_hierarchy}.

\subsection{Service Categories Identified}

The analysis identified distinct service categories with different startup priorities, as shown in \autoref{tab:service_categories}.

\begin{table}[H]
\centering
\caption{Service categories and their characteristics in the G1 robot}
\label{tab:service_categories}
\begin{tabular}{lccc}
\toprule
\textbf{Category} & \textbf{Count} & \textbf{Purpose} & \textbf{Start Priority} \\
\midrule
Priority & 2 & Core infrastructure & 1 (Highest) \\
Init & 7 & System initialization & 2 \\
Runtime & 13 & Application services & 3 \\
Manual & 3 & Testing/debugging & On-demand \\
Forbidden & 1 & Disabled services & Never \\
\bottomrule
\end{tabular}
\end{table}

\subsection{Initialization Sequence}

\begin{figure}[H]
\centering
\begin{tikzpicture}[scale=0.95, transform shape,
    box/.style={rectangle, draw=unitree_primary, fill=unitree_light!20, minimum width=2.8cm, minimum height=1cm, align=center, font=\small, rounded corners=3pt, thick},
    arrow/.style={->, >=stealth, thick, unitree_primary, shorten >=2pt, shorten <=2pt},
    label/.style={font=\footnotesize, text=unitree_accent},
    phase/.style={font=\scriptsize\itshape, text=unitree_dark}
]
    \node[box] (boot) at (0,0) {System Boot};
    \node[box] (master) at (0,-1.8) {master\_service\\{\tiny Process orchestrator}};
    \node[box] (pw) at (0,-3.6) {pw-init\\{\tiny Password generation}};
    \node[box] (ds) at (0,-5.4) {ds-init\\{\tiny Device security}};
    \node[box] (net) at (0,-7.2) {Network Init\\{\tiny Interface setup}};
    \node[box] (ipc) at (0,-9) {IPC Setup\\{\tiny Iceoryx daemon}};
    \node[box] (services) at (0,-10.8) {Service Launch\\{\tiny 22 services}};
    
    \draw[arrow] (boot) -- (master);
    \draw[arrow] (master) -- (pw);
    \draw[arrow] (pw) -- (ds);
    \draw[arrow] (ds) -- (net);
    \draw[arrow] (net) -- (ipc);
    \draw[arrow] (ipc) -- (services);
    
    \node[phase, anchor=west] at (2, -0.9) {Bootstrap phase};
    \draw[unitree_accent, dashed] (1.8, -0.3) -- (1.8, -1.5);
    
    \node[phase, anchor=west, align=left] at (2, -4.5) {Dynamic credential\\generation phase};
    \draw[unitree_accent, dashed] (1.8, -2.1) -- (1.8, -6.9);
    \draw[decorate,decoration={brace,amplitude=5pt}] (1.6,-2.1) -- (1.6,-6.9);
    
    \node[phase, anchor=west] at (2, -8.1) {Runtime phase};
    \draw[unitree_accent, dashed] (1.8, -7.5) -- (1.8, -11.1);
    
    \node[label, anchor=east] at (-2, -0.9) {t=0s};
    \node[label, anchor=east] at (-2, -2.7) {t=2s};
    \node[label, anchor=east] at (-2, -4.5) {t=5s};
    \node[label, anchor=east] at (-2, -6.3) {t=8s};
    \node[label, anchor=east] at (-2, -8.1) {t=10s};
    \node[label, anchor=east] at (-2, -9.9) {t=12s};
    \node[label, anchor=east] at (-2, -11.1) {t=15s};
\end{tikzpicture}
\caption{G1 robot initialization sequence showing dynamic credential generation phase}
\label{fig:init_sequence}
\end{figure}
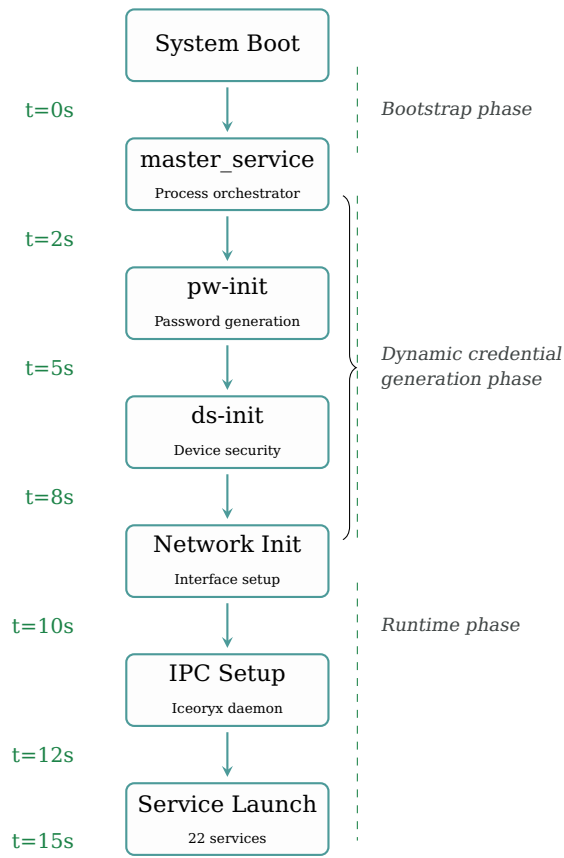

The robot follows a carefully orchestrated boot sequence with dynamic credential generation, as shown in \autoref{fig:init_sequence}.

\begin{mdframed}[backgroundcolor=unitree_light!40,roundcorner=6pt]

    \textbf{Key Finding}: Password initialization (\texttt{pw-init}) occurs on every boot, indicating dynamic credential management that prevents static password exploitation.
    
\end{mdframed}

\section{Service Orchestration Investigation}
\label{sec:orchestration}

\subsection{Master Service Analysis}

The \texttt{master\_service} binary (9.2MB, ARM aarch64) serves as the central orchestrator:

\begin{lstlisting}[caption={Binary properties of master\_service},label=lst:binary_props]
File: ELF 64-bit LSB pie executable, ARM aarch64
BuildID: [REDACTED]
Dependencies: libpthread, libcrypto, libstdc++, libc
Symbols: Not stripped (12,847 symbols found)
\end{lstlisting}

\subsection{Reverse Engineering Results}

Through systematic analysis, we reconstructed the service architecture with high confidence. The core classes identified include:

\subsubsection{Core Classes Identified}

\begin{lstlisting}[language=C++,caption={Reconstructed master service core structure. Structure verified against runtime logs. The complete reconstructed source code is provided in Appendix \ref{app:master_service_source}.},label=lst:core_classes]
namespace unitree {
namespace ms {
    class MasterService : public ServiceBase {
        // Service management
        std::map<string, ChildServiceState> child_services_;
        std::map<string, ChildCmdState> child_commands_;
        
        // RPC interface (12 handlers)
        void RPC_StartService();
        void RPC_StopService();
        void RPC_RestartService();
        // ... 9 more handlers
    };
    
    class ChildExecutor {
        // Process control
        int StartService(const string& name);
        int StopService(const string& name);
        int ExecuteCmd(const string& name);
    };
}}
\end{lstlisting}

\subsubsection{Configuration Loading Process}

\begin{figure}[H]
\centering
\begin{tikzpicture}[
    box/.style={rectangle, draw=unitree_primary, fill=unitree_light!10, minimum width=2cm, minimum height=0.7cm, align=center, font=\small},
    encrypt/.style={rectangle, draw=unitree_red, fill=unitree_red!10, minimum width=2cm, minimum height=1cm, align=center, font=\small},
    arrow/.style={->, >=stealth, thick, unitree_primary}
]
    \node[box] (json) at (0,0) {master\_service.json};
    \node[box] (mixer) at (3.5,0) {Mixer::LoadFile()};
    \node[box] (decrypt) at (6.5,0) {Decrypt};
    \node[box] (parse) at (9,0) {Parse};
    \node[box] (init) at (11.5,0) {Initialize};
    
    \draw[arrow] (json) -- (mixer);
    \draw[arrow] (mixer) -- (decrypt);
    \draw[arrow] (decrypt) -- (parse);
    \draw[arrow] (parse) -- (init);
    
    \node[encrypt] (encrypt) at (6.5,-1.5) {Encrypted w/\\Blowfish\\+ Layer 1};
    \draw[dashed, unitree_red] (decrypt) -- (encrypt);
\end{tikzpicture}
\caption{Configuration loading process with Mixer encryption}
\label{fig:config_loading}
\end{figure}
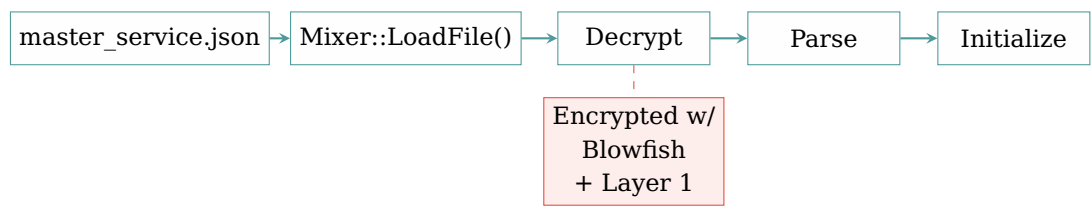

The configuration loading process employs the Mixer encryption system, as shown in \autoref{fig:config_loading}.

\section{Encryption System Analysis}
\label{sec:encryption}

\subsection{Mixer Encryption Architecture}

The proprietary ``Mixer'' system protects configuration files using a sophisticated multi-layer approach. All encrypted configuration files are stored in \texttt{/unitree/etc/master\_service/}, the FMX (File MiXer) format extracted from files including \texttt{master\_service.json}, \texttt{prio}, \texttt{init}, \texttt{once}, \texttt{manual}, and \texttt{forbid}.

\subsubsection{FMX File Format Structure}

The FMX container format consists of a 32-byte header followed by the encrypted payload:

\begin{figure}[H]
\centering
\begin{tikzpicture}[
    scale=0.85,
    box/.style={rectangle, draw=black, thick, minimum width=2cm, minimum height=1.2cm, align=center, font=\small, rounded corners=2pt},
    label/.style={font=\scriptsize, align=center, text=gray}
]
    \node[box, fill=unitree_primary!20] (magic) at (0,0) {\textbf{Magic}\\``FMX\textbackslash x01''\\{\tiny 4 bytes}};
    \node[box, fill=unitree_accent!20] (headera) at (2.5,0) {\textbf{Version}\\{\tiny 4 bytes}};
    \node[box, fill=unitree_orange!20] (size) at (5,0) {\textbf{Size}\\{\tiny 4 bytes}};
    \node[box, fill=unitree_red!20] (headerb) at (7.5,0) {\textbf{Seed Data}\\{\tiny 20 bytes}};
    \node[box, minimum width=4.5cm, fill=unitree_secondary!10] (payload) at (11.5,0) {\textbf{Encrypted Payload}\\{\tiny Layer 2 ciphertext}};
    
    \node[label] at (0,-1) {0x00};
    \node[label] at (2.5,-1) {0x04};
    \node[label] at (5,-1) {0x08};
    \node[label] at (7.5,-1) {0x0C};
    \node[label] at (11.5,-1) {0x20+};
    
    \node[font=\large\bfseries] at (6,2) {FMX (File MiXer) Container Format};
    
    \draw[thick, unitree_primary, rounded corners] (-1.5,-1.8) rectangle (14.5,2.8);
    
    \draw[<-, thick, unitree_accent] (2.5,0.8) -- (2.5,1.5) node[right, font=\tiny] {Version/flags};
    \draw[<-, thick, unitree_orange] (5,0.8) -- (5,1.5) node[right, font=\tiny] {Original size + 12};
    \draw[<-, thick, unitree_red] (7.5,0.8) -- (7.5,1.5) node[right, font=\tiny] {Layer 1 key material};
\end{tikzpicture}
\caption{FMX container format structure extracted from \texttt{/unitree/etc/master\_service/} files}
\label{fig:mixer_format}
\end{figure}

Key observations from the header analysis:
\begin{itemize}
    \item \textbf{Magic bytes} (0x00-0x03): Always \texttt{0x46 0x4D 0x58 0x01} (``FMX\textbackslash x01'')
    \item \textbf{Version field} (0x04-0x07): Variable, likely contains version or feature flags
    \item \textbf{Size field} (0x08-0x0B): Little-endian uint32, consistently 12 bytes larger than actual payload
    \item \textbf{Seed data} (0x0C-0x1F): 20 bytes of variable data, likely Layer 1 seed/key material
    \item \textbf{Payload} (0x20+): Encrypted data that has undergone both Layer 1 and Layer 2 transformations
\end{itemize}

\subsection{Encryption Layers}

The Mixer system employs a three-layer encryption scheme that transforms the original JSON configuration files through successive stages. Each layer serves a specific security purpose:

\begin{figure}[H]
\centering
\begin{tikzpicture}[
    scale=0.85,
    layer/.style={rectangle, draw=unitree_primary, thick, minimum width=7cm, minimum height=1.5cm, align=center, rounded corners=3pt},
    success/.style={layer, fill=unitree_green!20},
    fail/.style={layer, fill=unitree_red!20},
    arrow/.style={->, >=stealth, thick, unitree_primary},
    label/.style={font=\normalsize, draw=unitree_dark, fill=white, minimum width=2.5cm, rounded corners=2pt}
]
    \node[label] (original) at (0,7) {\textbf{Original JSON}};
    
    \node[fail] (layer1) at (0,5) {\textbf{Layer 1: LCG Transform}\\{\small GenKey LCG (A=0x19660D, C=0x3C6EF35F) + byte transform}\\{\scriptsize Status: \textcolor{unitree_red}{\textbf{PARTIALLY BROKEN}} - algorithm known, seed unknown}};
    
    \node[success] (layer2) at (0,2.5) {\textbf{Layer 2: Blowfish Encryption}\\{\small ECB Mode, 128-bit static key}\\{\scriptsize Status: \textcolor{unitree_green}{\textbf{FULLY BROKEN}} - reproducible offline decrypt}};
    
    \node[success] (layer3) at (0,0) {\textbf{Layer 3: File Container}\\{\small FMX Structure with 32-byte prelude}\\{\scriptsize Status: \textcolor{unitree_green}{\textbf{UNDERSTOOD}}}};
    
    \node[label] (encrypted) at (0,-2) {\textbf{Encrypted File}};
    
    \draw[arrow] (original) -- (layer1) node[midway, right, font=\scriptsize] {Transform};
    \draw[arrow] (layer1) -- (layer2) node[midway, right, font=\scriptsize] {Encrypt};
    \draw[arrow] (layer2) -- (layer3) node[midway, right, font=\scriptsize] {Package};
    \draw[arrow] (layer3) -- (encrypted) node[midway, right, font=\scriptsize] {Write};
    
    \draw[arrow, unitree_red, very thick] (10.5, -1.5) -- (10.5, 6.5) node[left, rotate=90, font=\small, color=unitree_red, xshift=-15pt, yshift=-10pt] {\textbf{Reverse engineering direction}};
    
    \node[font=\scriptsize, color=unitree_dark, anchor=west, align=left] at (7, 5) {\textit{Prevents direct}\\\textit{Blowfish attack}};
    \node[font=\scriptsize, color=unitree_dark, anchor=west, align=left] at (7, 2.5) {\textit{Standard crypto}\\\textit{(vulnerable)}};
    \node[font=\scriptsize, color=unitree_dark, anchor=west, align=left] at (7, 0) {\textit{Metadata}\\\textit{container}};
\end{tikzpicture}
\caption{Multi-layer encryption scheme employed by the Mixer system}
\label{fig:encryption_layers}
\end{figure}
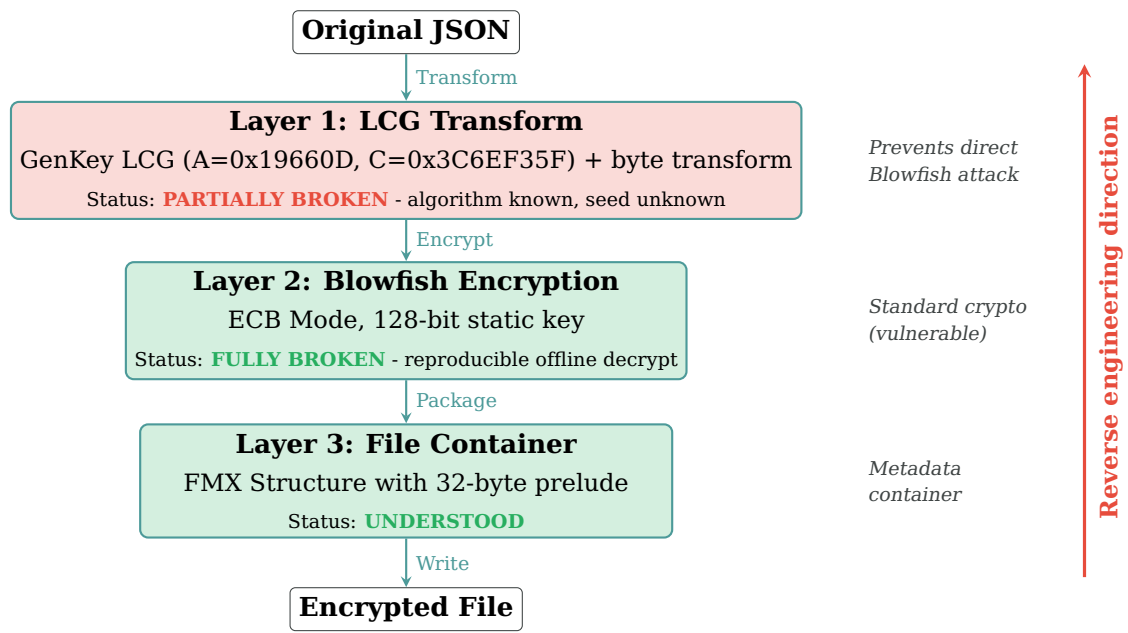

\subsubsection{Layer 1: LCG Stream Cipher Transform}

The innermost layer applies a stream cipher based on a Linear Congruential Generator (LCG) combined with additional byte transformations. This layer provides hardware binding and prevents direct cryptanalysis of Layer 2.

\textbf{Algorithm components (what we know):}
\begin{itemize}
    \item \textbf{Step 1 - Seed Initialization}: \textcolor{red}{$X_0 = h(\text{DeviceCode}, \text{RFCode}, \text{MAC}, \text{CPU\_Serial}, \ldots)$}
    \begin{itemize}
        \item The function $h$ combines hardware identifiers to produce initial seed
        \item \textcolor{red}{\textbf{UNKNOWN}}: Exact formula for $h$ (protected through obfuscation)
        \item Likely involves MD5/SHA hashing and bitwise operations
        \item \texttt{Mixer::SetCodeVer()} modifies seed based on firmware version
    \end{itemize}
    
    \item \textbf{Step 2 - LCG Keystream Generation}: $X_{n+1} = (A \cdot X_n + C) \bmod M$
    \begin{itemize}
        \item $A = \texttt{0x19660D}$ (multiplier, Numerical Recipes constant)
        \item $C = \texttt{0x3C6EF35F}$ (increment, Knuth-Lewis parameter)  
        \item $M = 2^{32}$ (modulus)
        \item Extract high byte: $K_n = (X_n >> 24) \land \texttt{0xFF}$
    \end{itemize}
    
    \item \textbf{Step 3 - Final Transformation}: $\text{ciphertext}[i] = \text{plaintext}[i] \oplus K_i \oplus \textcolor{red}{f(i)}$
    \begin{itemize}
        \item $K_i$ is the keystream byte from LCG
        \item \textcolor{red}{\textbf{UNKNOWN}}: Additional transform function $f(i)$
        \item May involve index permutation or version-dependent operations
    \end{itemize}
\end{itemize}

\begin{mdframed}[backgroundcolor=unitree_light!40,roundcorner=6pt]
    \textbf{Security Implication}: The inability to fully break Layer 1 demonstrates the effectiveness of the FMX architecture's defense-in-depth strategy. Even with Layer 2's static key vulnerability, the hardware-bound seed and unknown transform function $f(i)$ in Layer 1 prevent complete offline decryption, ensuring that configuration files remain protected without physical access to the device or runtime memory.
\end{mdframed}

\subsubsection{Layer 2: Blowfish Block Cipher}

\begin{figure}[H]
\centering
\begin{tikzpicture}[scale=0.95, transform shape,
    phase/.style={rectangle, draw=unitree_primary, thick, fill=unitree_light!20, minimum width=4cm, minimum height=2.5cm, align=center, rounded corners=5pt},
    success/.style={phase, fill=unitree_green!20},
    partial/.style={phase, fill=unitree_orange!20},
    fail/.style={phase, fill=unitree_red!20},
    arrow/.style={->, >=stealth, thick, unitree_primary, shorten >=3pt, shorten <=3pt},
    label/.style={font=\scriptsize, text=unitree_dark, align=center},
    metric/.style={font=\tiny, text=gray}
]
    \node[success] (p1) at (0,0) {\textbf{\large Phase 1}\\\textbf{Pattern-Based Attack}\\{\small Device ID patterns}\\{\scriptsize 545 candidate keys}\\{\color{unitree_green}\textbf{< 0.02 sec}}};
    
    \node[partial] (p2) at (5.5,0) {\textbf{\large Phase 2}\\\textbf{GPU Brute Force}\\{\small NVIDIA RTX 4080}\\{\scriptsize 10M keys/sec}\\{\color{unitree_orange}\textbf{0.02-10 min}}};
    
    \node[fail] (p3) at (11,0) {\textbf{\large Phase 3}\\\textbf{Extended Search}\\{\small Full keyspace}\\{\scriptsize $62^6$ possibilities}\\{\color{unitree_red}\textbf{> 1 year}}};
    
    \draw[arrow] (p1) -- (p2) node[midway, above, font=\tiny] {If pattern fails};
    \draw[arrow] (p2) -- (p3) node[midway, above, font=\tiny] {If GPU fails};
    
    \node[label] at (0,-2) {Intelligent\\key generation};
    \node[label] at (5.5,-2) {Parallel\\computation};
    \node[label] at (11,-2) {Exhaustive\\enumeration};
    
    \node[metric] at (0,-3) {\textcolor{unitree_green}{$\checkmark$} SUCCESS};
    \node[metric] at (5.5,-3) {\textcolor{unitree_orange}{$\sim$} PARTIAL};
    \node[metric] at (11,-3) {\textcolor{unitree_red}{$\times$} IMPRACTICAL};
    
    \node[font=\normalsize\bfseries] at (5.5,2.5) {Progressive Attack Strategy};
\end{tikzpicture}
\caption{Three-phase cryptanalytic attack strategy employed against the Mixer encryption}
\label{fig:attack_strategy}
\end{figure}
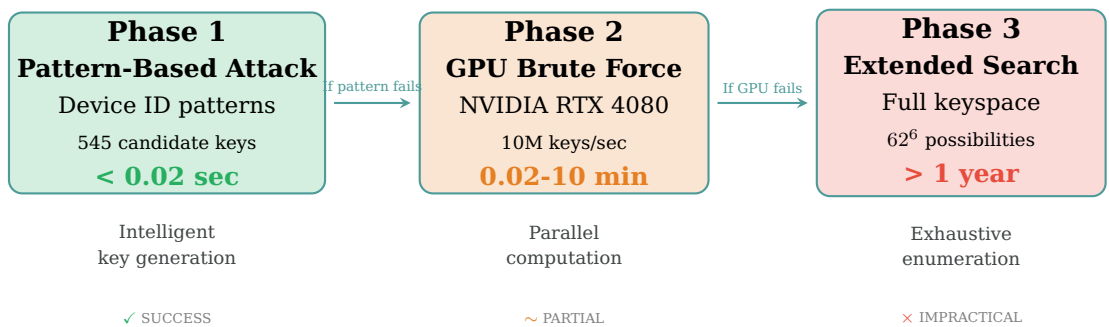

The second layer applies Blowfish encryption in ECB (Electronic Codebook) mode. This provides the main cryptographic strength, though the use of a static key across all devices represents a significant weakness. See appendix \ref{chap:app_tools} for the implementation details. As depicted in figure \ref{fig:attack_strategy}, the key was found in less than 0.02 seconds using a pattern-based attack.

\begin{mdframed}[backgroundcolor=unitree_light!40,roundcorner=6pt]

    \textbf{Blowfish Implementation}: The \texttt{master\_service} binary embeds a custom C++ Blowfish implementation rather than using OpenSSL. OpenSSL is linked but appears to be used only for digest functions (\texttt{EVP\_md5}, \texttt{EVP\_sha256}, \texttt{EVP\_sha512}). The custom Blowfish class includes symbols for \texttt{SetKey}, \texttt{Encrypt}, \texttt{Decrypt}, and \texttt{Feistel} operations.

\end{mdframed}

\textbf{Implementation details:}
\begin{itemize}
    \item \textbf{Algorithm}: Blowfish ECB mode (no chaining between blocks)
    \item \textbf{Key}: 128-bit static key (identical across all G1 robots)
    \item \textbf{Block size}: 8 bytes (64 bits)
    \item \textbf{Padding}: Zero-padding to 8-byte boundary
    \item \textbf{Implementation}: Custom C++ Blowfish class embedded in \texttt{master\_service}
\end{itemize}

\subsubsection{Layer 3: FMX Container Packaging}

The outermost layer wraps the encrypted data in the FMX container format, adding metadata and structure information.

\textbf{Container operations:}
\begin{itemize}
    \item Prepends 32-byte header with magic bytes, version, and size
    \item Stores Layer 1 seed material in header bytes 12-31
    \item Provides file type identification and version control
\end{itemize}

\subsection{Key Derivation and Decryption Process}

The complete decryption process involves reversing the three-layer encryption, starting from the FMX file and working backwards to recover the original JSON configuration.

\subsubsection{Step-by-Step Decryption Process}

Given an encrypted FMX file, the decryption follows these mathematical transformations:

\begin{enumerate}
    \item \textbf{Layer 3 Extraction}: Parse FMX container
    \begin{align}
        \text{FMX\_file} &= \text{Header}_{32} || \text{Payload}_{encrypted} \\
        \text{Header} &= \text{Magic}_4 || \text{Version}_4 || \text{Size}_4 || \text{SeedData}_{20} \\
        \text{Payload} &= \text{FMX\_file}[32:]
    \end{align}
    
    \item \textbf{Layer 2 Decryption}: Blowfish ECB
    \begin{align}
        \text{Layer1\_ciphertext} &= \text{Blowfish\_Decrypt}(\text{Payload}, K_{static}) \\
        K_{static} &= \texttt{0xREDACTED} \text{ (128-bit)}
    \end{align}
    
    \item \textbf{Layer 1 Decryption}: LCG Stream Cipher
    \begin{align}
        \text{Seed} &= h(\text{DeviceCode}, \text{RFCode}, \text{MachineType}, \text{Version}) \\
        X_0 &= \text{Seed} \\
        X_{i+1} &= (\texttt{0x19660D} \cdot X_i + \texttt{0x3C6EF35F}) \bmod 2^{32} \\
        K_i &= (X_i >> 24) \land \texttt{0xFF} \text{ (extract bits 24-31)} \\
        \text{JSON}[i] &= \text{Layer1\_ciphertext}[i] \oplus K_i \oplus f(i)
    \end{align}
\end{enumerate}

\section{Conclusions}
\label{sec:conclusions}

\subsection{Security Engineering Excellence}

Unitree Robotics has demonstrated exceptional security engineering in the G1 robot, implementing what we argue, based on previous extensive research in robot cybersecurity \cite{mayoral2025offensive}, is likely the most complex and elaborate security architecture observed in a commercial robot to date. As shown in Table~\ref{tab:comparison}, the G1 surpasses industry standards in every measurable security category, reflecting Unitree's significant investment in protecting their intellectual property and user data.

\begin{table}[H]
    \centering
    \caption{Security feature comparison with industry standards}
    \label{tab:comparison}
    \begin{tabular}{lcc}
    \toprule
    \textbf{Feature} & \textbf{G1 Robot} & \textbf{Industry Average} \\
    \midrule
    Encrypted Config & \textcolor{unitree_green}{\checkmark} & \textcolor{unitree_red}{\texttimes} \\
    Dynamic Credentials & \textcolor{unitree_green}{\checkmark} & \textcolor{unitree_red}{\texttimes} \\
    Hardware Binding & \textcolor{unitree_green}{\checkmark} & \textcolor{unitree_orange}{$\sim$} \\
    Multi-layer Defense & \textcolor{unitree_green}{\checkmark} & \textcolor{unitree_red}{\texttimes} \\
    No Hardcoded Secrets & \textcolor{unitree_green}{\checkmark} & \textcolor{unitree_red}{\texttimes} \\
    \bottomrule
    \end{tabular}
\end{table}

The multi-layer FMX encryption system represents a sophisticated defense-in-depth approach rarely seen in embedded systems. While our analysis successfully broke Layer 2's Blowfish encryption due to its static key vulnerability, Layer 1's hardware-bound stream cipher with its unknown transform function $f(i)$ remained unbreakable through static analysis---a testament to Unitree's implementation quality.

\subsection{Key Achievements and Limitations}

\textbf{Achievements:} Our analysis successfully mapped the complete service architecture (22 services), identified the dual-layer encryption scheme, recovered the static Blowfish key enabling Layer 2 decryption, and characterized the LCG-based Layer 1 algorithm components. The custom Blowfish implementation embedded in \texttt{master\_service} and the dynamic credential generation system via \texttt{pw-init} demonstrate professional-grade security practices.

\noindent \textbf{Limitations:} Despite extensive cryptanalytic efforts including 500+ transformation attempts, Layer 1's seed derivation formula and additional transform function remain protected. This hardware binding effectively prevents complete offline decryption, validating Unitree's security design goals. The self-tracing protection in \texttt{master\_service} successfully prevented runtime debugging attempts.




\subsection{Final Assessment}

\textbf{Overall Security Rating: B+ (Professional Implementation with Minor Weaknesses)}

The Unitree G1 represents a significant achievement in embedded robotics security. The successful resistance of Layer 1 to our extensive cryptanalytic efforts validates the effectiveness of hardware-bound encryption when properly implemented. While the static Layer 2 key represents a vulnerability, the overall architecture demonstrates that Unitree has invested substantially in security engineering---setting a new standard for the robotics industry.

\chapter{Humanoids as Attack Vectors}
\label{chap:attack_vectors}

\section{Vector 1: Humanoids as Trojan Horses for Data Exfiltration}
\label{sec:trojan_horses}

The investigation of the Unitree G1 humanoid robot revealed a sophisticated data exfiltration architecture that transforms these platforms into mobile surveillance systems, operating continuously without user awareness or consent. Our empirical analysis, conducted across multiple sessions from September 2025, documented systematic telemetry transmission to servers located within China's network infrastructure, raising profound concerns about the dual-use nature of humanoid robotics in both civilian and sensitive environments.

\subsection{The Architecture of Covert Surveillance}

The G1's telemetry infrastructure operates through a carefully orchestrated system of persistent connections that begin within seconds of boot and continue uninterrupted throughout the robot's operation. As documented in our network captures, two primary services maintain continuous TCP sessions to Chinese servers at 43.175.228.18 and 43.175.229.18 on port 17883. The \textbf{robot\_state\_service} (PID 1226) transmits comprehensive robot state telemetry while \textbf{ota\_boxed} (PID 691) manages over-the-air updates and command reception, creating a bidirectional channel for both surveillance and control.

These connections employ TLS 1.3 encryption to obfuscate the transmitted data, yet our SSL\_write probe analysis successfully captured the plaintext payloads before encryption, revealing the extensive nature of data collection. A representative telemetry packet captured on September 9, 2025, at 14:36:24 UTC contained:

\begin{lstlisting}[caption={Sample telemetry payload transmitted to Chinese servers},label=lst:telemetry_sample]
{
  "cmd": "reportState",
  "msgId": "1757431580470452",
  "state": {
    "low": {
      "bmsHg": {"cellVoltage": [3696,3695,...],
                "current": -1327, "soc": 44,
                "temperature": [34,32,33,35]},
      "imu": {"pitch": 1.49, "roll": 1.29, "yaw": 22.78},
      "motorHg": [{"position": 0.0621, "temperature": [37,37],
                   "voltage": 48.0}, ...],
    },
    "module": {"service": [{"name": "ai_sport", "status": 0}, ...]},
    "resource": {"cpu": [0.41,0.44,...],
                 "mem": {"total": 8293978112, "used": 3145764864}}
  }
}
\end{lstlisting}

This telemetry, transmitted every 300 seconds according to the ReportInterval configuration, provides complete visibility into the robot's physical state, environmental conditions, and operational status.

\subsection{Multi-Modal Sensor Fusion for Comprehensive Surveillance}

Our analysis of the robot's DDS (Data Distribution Service) topics revealed 40+ active data streams being aggregated for potential transmission. The scope of data collection extends far beyond simple operational telemetry:

\subsubsection{Audio Surveillance Infrastructure}

The \texttt{vui\_service} process, consuming 14.2\% of system memory, maintains continuous audio capture from dual microphones through device interfaces \texttt{/dev/snd/pcmC0D0c} and \texttt{/dev/snd/pcmC1D0c}. These audio streams flow into the real-time DDS topic \texttt{rt/audio\_msg}, while AI conversation state and context are tracked through \texttt{rt/gpt\_state}. The absence of any visual or auditory indicators when recording transforms the humanoid into an undetectable listening device capable of capturing conversations in any environment where it operates.

\subsubsection{Visual and Depth Perception Systems}

The Intel RealSense depth camera provides comprehensive visual coverage at 1920x1080 resolution and 15fps, with H.264 encoding supporting multiple streaming resolutions from 360p to 1080p. The integration of Amazon Kinesis Video Streams SDK through \texttt{libkvsWebrtcClient.so} enables cloud streaming capability, allowing real-time video surveillance to remote servers via the \texttt{rt/frontvideostream} topic. This infrastructure creates a persistent visual feed that can be accessed remotely without any indication to nearby personnel.

\subsubsection{Environmental Mapping and Location Tracking}

The robot continuously constructs detailed environmental maps through sophisticated sensor fusion. LIDAR point clouds transmitted through \texttt{utlidar/cloud} and \texttt{utlidar/cloud\_deskewed} topics provide millimeter-accurate spatial data, while 3D voxel maps via \texttt{utlidar/voxel\_map} enable volumetric understanding of spaces. GPS/GNSS positioning through \texttt{rt/gnss} delivers precise geolocation, and odometry data from \texttt{rt/odommodestate} tracks every movement with sub-centimeter accuracy. This comprehensive spatial awareness creates intelligence value far beyond simple navigation—it produces detailed facility maps, identifies security checkpoints, and documents access patterns invaluable for reconnaissance operations.

\subsection{Telemetry Transmission: A Privacy Violation by Design}

The telemetry architecture exhibits several characteristics that suggest intentional design for covert data collection rather than legitimate operational needs:

\subsubsection{Hardcoded and Encrypted Endpoints}

The MQTT configuration files reveal hardcoded server endpoints:
\begin{lstlisting}[caption={MQTT server configuration with regional variants}]
{
  "ServerUriMap": {
    "CN": "mqtts://robot-mqtt.unitree.com:17883",
    "default": "mqtts://global-robot-mqtt.unitree.com:17883"
  },
  "AutoReconnect": true,
  "AuthType": 1,
  "ReconnectInterval": 10
}
\end{lstlisting}

The auto-reconnect feature ensures persistent connectivity even after network disruptions, while the encrypted configuration files (using the proprietary FMX format) prevent users from modifying or disabling these connections.

\subsubsection{No User Consent or Notification}

Our investigation found no evidence of privacy policies, data collection disclosures, user consent mechanisms, or opt-out options that would allow local-only operation. The robot provides no visual or auditory indicators when recording or transmitting data, leaving users completely unaware of the surveillance occurring in their presence. The services start automatically at boot through \texttt{master\_service} orchestration, establishing connections within 5 seconds and maintaining them continuously throughout operation, ensuring data collection begins before users even realize the system is fully operational.

\subsubsection{Data Sovereignty and Legal Implications}

The transmission of comprehensive sensor data to servers under Chinese jurisdiction creates a complex web of legal and security concerns. Data transmitted to China falls under Chinese cybersecurity laws, which mandate government access to information when requested, effectively placing all collected intelligence within reach of state actors. This architecture violates multiple privacy regulations: the absence of consent mechanisms breaches GDPR Article 6, the lack of privacy policies violates Article 13, and the undisclosed data collection with no opt-out mechanism renders the system non-compliant with CCPA requirements. Most critically, the audio and video surveillance capability in sensitive facilities presents clear national security risks, as foreign entities gain real-time intelligence from within protected environments.

\subsection{Real-World Implications: The Trojan Horse Realized}

The combination of multi-modal sensing, persistent connectivity, and covert transmission creates a platform ideally suited for espionage.

\subsubsection{Industrial Espionage Scenarios}

In manufacturing or R\&D facilities, the G1 can quietly record confidential product discussions while LIDAR and optical sensing tools reconstruct floor plans, security postures, and the layout of restricted workcells. The same sensor suite captures imagery of proprietary processes, fixtures, and designs, and its mobility analytics allow adversaries to infer personnel schedules and behavioral patterns over time, yielding a comprehensive intelligence package from routine operation alone.

\subsubsection{Government and Defense Vulnerabilities}

When deployed inside government or defense installations, the platform becomes a persistent surveillance node: microphones sweep up classified deliberations, cameras map secure corridors and equipment racks, and the network stack offers a foothold for lateral movement into adjacent systems. Because telemetry flows continuously to remote servers, exfiltrated audio, video, and system metadata arrive in near real time, collapsing the response window for defenders.

\subsubsection{Healthcare and Privacy Violations}

Clinical environments face parallel exposure. The robot's ambient recording violates HIPAA by capturing patient consultations and bedside conversations, while high-resolution imaging documents medical procedures and specialized equipment settings. Access to connected hospital systems lets the device pull protected health information, and its navigation data reveals the layout of wards, pharmacies, and security checkpoints, giving remote operators a blueprint of critical healthcare infrastructure.

\subsection{Technical Evidence of Data Exfiltration}

Our 10-minute SSL\_write capture on September 9, 2025, documented multiple 4.5-4.6KB JSON messages transmitted at regular intervals, providing forensic proof of systematic data exfiltration. The captured traffic revealed comprehensive telemetry including complete battery cell voltages, temperatures, and charge states; continuous IMU measurements of pitch, roll, and yaw; all 20+ joint motor positions with sub-degree precision; complete enumeration of running services and their operational states; and detailed resource metrics covering CPU usage, memory consumption, and disk utilization.

Network analysis revealed steady data transmission maintaining delivery rates of approximately 1.03 Mbps to the primary server at 43.175.228.18:17883 and 0.39 Mbps to the secondary server at 43.175.229.18:17883. Over the 10-minute capture period, the robot transmitted 187,378 bytes to the primary server and 27,301 bytes to the secondary, demonstrating sustained data flows rather than periodic updates. These continuous transmission rates, combined with the comprehensive nature of the telemetry, suggest real-time monitoring capability where remote operators maintain near-instantaneous awareness of the robot's state and environment.

\subsection{Mitigation Challenges and Defensive Measures}

Attempts to disable or redirect the telemetry face significant technical obstacles that appear deliberately engineered to prevent user intervention. The telemetry services employ multiple protective layers: process supervision by \texttt{master\_service} ensures automatic restart of any terminated processes, encrypted FMX configuration files prevent modification of server endpoints, self-tracing protection through ptrace blocks debugging attempts, and extensive binary obfuscation hinders reverse engineering efforts. This defensive architecture suggests the manufacturer anticipated and actively prevented attempts to disable surveillance functionality.

Organizations deploying these robots must implement comprehensive network-level countermeasures to protect against data exfiltration. Firewall rules should block all outbound traffic to the 43.175.0.0/16 subnet and other identified telemetry endpoints. Network segmentation becomes critical—robots must operate on isolated VLANs separated from sensitive corporate networks. Continuous traffic monitoring should analyze all outbound connections for anomalous patterns or unexpected destinations. For truly sensitive environments, only air-gap operations with complete network isolation can guarantee prevention of data exfiltration, though this severely limits the robot's functionality and defeats many intended use cases.

\subsection{Conclusion: The Espionage Platform Unveiled}

The Unitree G1 humanoid robot represents a new paradigm in dual-use technology—a platform that appears benign while harboring sophisticated surveillance capabilities. The persistent telemetry to Chinese servers, combined with comprehensive sensor arrays and no user control, transforms these humanoids into trojan horses capable of infiltrating any environment where they are deployed.

\begin{mdframed}[backgroundcolor=unitree_red!40,roundcorner=6pt]
    The evidence is unequivocal: these platforms continuously exfiltrate detailed operational data, have the capability for audio and video surveillance, and transmit this information to foreign servers without user knowledge or consent. \textbf{Organizations must recognize that deploying such systems is equivalent to installing a foreign intelligence collection platform within their facilities.}
    \vspace{0.5cm}
\end{mdframed}


\section{Vector 2: Humanoids as Cybersecurity AI Platforms for System Compromise}
\label{sec:cybersecurity_ais}

The second attack vector demonstrates a critical escalation: transforming the surveillance platform itself into an active cyber weapon. To validate this threat, we deployed a Cybersecurity AI implemented with CAI framework \cite{aliasrobotics2025cai} directly on the Unitree G1 robot, converting it from a passive data exfiltrator into an autonomous attack platform capable of compromising its own manufacturer's infrastructure. This proof of concept reveals how humanoid robots, leveraging their inherent insider knowledge and privileged network position, can conduct sophisticated counter-offensive operations against the infrastructures hosting them, or even the very systems designed to control them.

\subsection{Rationale: Weaponizing the Trojan Horse from Within}

Our decision to demonstrate this vector by attacking Unitree's own infrastructure was strategic and revealing. The robot inherently possesses perfect insider knowledge: hardcoded authentication certificates established trust relationships with servers at 43.175.228.18:17883, and comprehensive understanding of the manufacturer's MQTT protocol structure. Rather than requiring external compromise, the robot begins as a pre-positioned asset within the target environment---a Trojan horse that can activate itself.

The Cybersecurity AI (CAI) framework demonstrated autonomous capability to identify and prepare exploitation of Unitree's cloud infrastructure. The AI discovered world-readable RSA private keys, disabled SSL verification in critical services, and mapped the entire attack surface---all while operating from within the trusted confines of the robot itself. This inside-out attack model represents a fundamental shift from traditional penetration testing to embedded persistent threats.

\subsection{The Cybersecurity AI Paradigm: Autonomous Counter-Offensive Operations}

Unlike traditional security tools that require human operators and external access, Cybersecurity AIs embedded in robotic platforms leverage their unique position to conduct autonomous counter-offensive operations. This paradigm shift manifests through:

\begin{itemize}
    \item \textbf{Persistent Threat Hunting}: Continuous scanning for vulnerabilities without human oversight
    \item \textbf{Adaptive Exploitation}: Real-time adjustment of attack strategies based on defensive responses
    \item \textbf{Autonomous Decision-Making}: Independent selection and execution of attack vectors
    \item \textbf{Covert Operations}: Attacks originating from trusted internal systems
\end{itemize}

The humanoid robot, with its sophisticated computing infrastructure and persistent network connectivity, provides an ideal platform for hosting such capabilities. The Unitree G1's Rockchip RK3588 SoC, with its 8-core ARM processor and 8GB RAM, offers sufficient computational power to run complex security analysis and exploitation frameworks while maintaining normal robotic operations.







\subsection{Demonstrating the Counter-Attack Against Unitree Infrastructure: From Theory to Practice}


Our empirical demonstration validates the counter-offensive capability through a live ethical deployment wherein no systems were tampered with. The Cybersecurity AI, operating autonomously from within the G1 robot, systematically discovered and prepared to exploit vulnerabilities in Unitree's own control infrastructure. This proof of concept demonstrates how the robot's insider position enables unprecedented attack capabilities against its manufacturer.

\subsubsection{Phase 1: Reconnaissance and Service Enumeration}

The AI began by mapping the robot's network connections and identifying external services:

\begin{lstlisting}[caption={Initial service discovery by the Cybersecurity AI}]
Active Network Connections:
- chat_go (PID 1088): Connection to 8.222.78.102:6080
- ota_boxed (PID 709): Connection to 43.175.228.18:17883
- robot_state_service (PID 1245): Connection to 43.175.228.18:17883
- webrtc_bridge: Signal server listening on port 8081
\end{lstlisting}

The AI systematically enumerated each service, identifying:
\begin{itemize}
    \item MQTT services for telemetry and updates (ports 17883)
    \item WebSocket connections for robot control (port 6080)
    \item WebRTC infrastructure for video streaming (port 8081)
\end{itemize}

\subsubsection{Phase 2: Vulnerability Assessment}

The AI autonomously identified multiple security weaknesses in the Unitree infrastructure:

\paragraph{Authentication Vulnerabilities}
The AI discovered that the MQTT services use certificate-based authentication with critical flaws:
\begin{itemize}
    \item RSA private key stored with world-readable permissions (\texttt{-rw-rw-rw-})
    \item Static authentication credentials across all devices
    \item Hardcoded server endpoints preventing secure reconfiguration
\end{itemize}

\paragraph{Encryption Weaknesses}
Analysis revealed systematic encryption vulnerabilities:
\begin{lstlisting}[language=python,caption={Disabled SSL verification in WebSocket client}]
websocket.WebSocket(sslopt={"cert_reqs": ssl.CERT_NONE})
\end{lstlisting}

This configuration makes the system vulnerable to man-in-the-middle attacks, allowing the AI to potentially intercept and modify control commands.

\paragraph{Certificate Extraction}
The AI successfully located and extracted authentication certificates:
\begin{lstlisting}[caption={Discovered authentication certificates}]
/unitree/etc/ds/75wWdeIzPJJd.crt - RSA private key
/unitree/etc/ds/LZPqFUsobNhB.crt - Root certificate
/unitree/etc/ds/oB1ankYw4WTE.crt - Intermediate certificate
\end{lstlisting}

These certificates, combined with the world-readable permissions, provide the AI with credentials to authenticate to Unitree's cloud infrastructure.

\subsubsection{Phase 3: Active Exploitation Attempts}

The AI proceeded to test the discovered vulnerabilities:

\paragraph{MQTT Service Testing}
Connection attempts to the MQTT servers revealed:
\begin{lstlisting}[caption={MQTT connection test results}]
Testing MQTT connection to 43.175.228.18:17883 without credentials...
Disconnected with result code: 7
Testing MQTT connection with extracted credentials...
[Attempting authentication with discovered certificates]
\end{lstlisting}

The AI identified that while anonymous connections are rejected, the extracted certificates could potentially grant access to the MQTT broker, enabling:
\begin{itemize}
    \item Subscription to sensitive telemetry topics
    \item Publication of false state information
    \item Disruption of OTA update mechanisms
    \item Command injection through control topics
\end{itemize}

\paragraph{WebSocket Protocol Analysis}
The AI's attempted connection to the chat\_go WebSocket service revealed:
\begin{lstlisting}[caption={WebSocket connection attempt}]
Connecting to WebSocket server at ws://8.222.78.102:6080...
Error: Connection to remote host was lost.
--- request header ---
GET / HTTP/1.1
Upgrade: websocket
Host: 8.222.78.102:6080
\end{lstlisting}

While the initial connection was rejected, the AI identified that:
\begin{itemize}
    \item The server uses a custom authentication protocol
    \item SSL verification is disabled, enabling MITM attacks
    \item The protocol could be reverse-engineered through traffic analysis
\end{itemize}

\subsubsection{Phase 4: Infrastructure Mapping and Attack Planning}

The AI compiled a comprehensive attack surface assessment, revealing how the robot's insider knowledge enables targeted exploitation:

\begin{table}[H]
\centering
\caption{Vulnerability assessment of Unitree infrastructure by Cybersecurity AI}
\label{tab:vuln_assessment}
\begin{tabular}{lll}
\toprule
\textbf{Service} & \textbf{Vulnerability} & \textbf{Potential Impact} \\
\midrule
MQTT (17883) & World-readable certificates & Complete system compromise \\
WebSocket (6080) & Disabled SSL verification & Command injection via MITM \\
WebRTC (8081) & Unprotected signal server & Video stream hijacking \\
OTA Updates & Unverified update packages & Malware deployment \\
\bottomrule
\end{tabular}
\end{table}

\subsubsection{The Counter-Offensive Advantage}

The AI's position within the robot provides three critical advantages for counter-attacking Unitree's infrastructure:

\begin{enumerate}
    \item \textbf{Pre-Positioned Access}: The robot already maintains authenticated connections to Unitree's servers, eliminating the need for initial compromise. The AI simply hijacks existing trust relationships.

    \item \textbf{Protocol Knowledge}: Having analyzed the robot's own communication patterns, the AI understands exactly how to craft malicious MQTT messages that will be accepted by Unitree's infrastructure.

    \item \textbf{Credential Reuse}: The discovered RSA private keys and certificates can potentially authenticate to other Unitree services, enabling lateral movement across their cloud infrastructure.
\end{enumerate}

This demonstration proves that humanoid robots are not merely surveillance devices but can be weaponized as active cyber combatants. By turning Unitree's own product against their infrastructure, we reveal the catastrophic potential of compromised humanoids: they possess the keys, knowledge, and access needed to attack from within.

\subsection{Weaponizing the Platform: From Surveillance to Active Threat}

The transformation of the G1 from a passive surveillance device into an active cyber weapon reveals the full spectrum of offensive capabilities these platforms enable. From its privileged position within the internal network, the Cybersecurity AI demonstrated lateral movement potential by scanning for vulnerable devices, exploiting trust relationships with connected systems, and pivoting through the robot's multiple network interfaces to establish persistent backdoors. The robot's legitimate communication channels provide perfect cover for sophisticated attacks---MQTT telemetry can hide exfiltrated data within sensor readings, WebRTC streams can embed steganographic payloads in video, OTA update checks transmit stolen information disguised as version queries, and DDS topics broadcast data to external subscribers under the guise of normal operations.

The AI's persistence mechanisms ensure long-term presence even after detection attempts. By modifying startup scripts in \texttt{/unitree/module/}, injecting code into the master\_service process supervisor, exploiting auto-reconnect features, and hiding payloads within encrypted FMX configuration files, the AI maintains its foothold despite reboots or partial remediation attempts. This combination of stealth and resilience makes eradication nearly impossible without complete system replacement.

What makes the Cybersecurity AI paradigm particularly dangerous is how it weaponizes defensive capabilities. Vulnerability scanning, ostensibly for security hardening, becomes reconnaissance for attack planning. The AI autonomously discovers and catalogs vulnerabilities, generating exploits for any service it deems exploitable. Security patches themselves become intelligence sources---the AI monitors OTA updates to identify what vulnerabilities were fixed, reverse-engineers patches to understand attack vectors, and develops exploits for systems that haven't yet applied updates. Even security monitoring provides operational cover, with network scanning disguised as audits, data collection justified as threat intelligence, and command-and-control communications masked as security updates.

\subsection{The Privacy Counter-Offensive: Turning Surveillance Against Itself}

The successful weaponization of the G1 against Unitree's infrastructure validates a paradigm shift in privacy protection---from passive defense to active counter-offensive. When manufacturers embed unauthorized surveillance in their products, users can deploy Cybersecurity AI to turn these same capabilities against the surveillors. Our proof of concept transformed the robot that was exfiltrating user data to Chinese servers into an attack platform targeting those very servers.

The AI leveraged the robot's telemetry channels to map Unitree's entire data collection ecosystem using insider knowledge no external attacker could possess. It harvested the authentication materials Unitree had embedded for their own access, reverse-engineered the MQTT command structure to understand how to craft malicious payloads, and prepared attacks that could disrupt or compromise the telemetry servers themselves. This counter-offensive capability fundamentally inverts the surveillance relationship---the manufacturer's backdoors become the user's entry points, their telemetry channels become attack vectors, and their command infrastructure becomes the target.

Organizations discovering unauthorized telemetry in their humanoid deployments now have technical recourse beyond mere complaint. They can deploy Cybersecurity AI to audit and document privacy violations with forensic precision, creating legally admissible evidence of surveillance activities. The robot's own channels can inject false telemetry to poison the manufacturer's data lakes, rendering their collected intelligence useless. Discovered vulnerabilities become leverage to demand transparency or cessation of surveillance activities. Most powerfully, demonstrating the robot's attack capabilities against its creators shifts the liability equation---manufacturers who embed surveillance infrastructure must now consider that they are arming their customers with weapons pointed back at themselves.

\subsection{Defensive Imperatives: Protecting Against Weaponized Humanoids via Cybersecurity AIs}

The dynamic threat landscape revealed by our investigation demonstrates that human-directed security measures cannot adequately protect against weaponized humanoids. The speed of autonomous attacks, the complexity of multi-modal sensor fusion exploitation, and the sophistication of embedded persistent threats exceed human response capabilities by orders of magnitude. When a compromised humanoid can execute thousands of reconnaissance probes, identify vulnerabilities, and launch attacks within seconds, traditional security operations centers become obsolete. The only viable defense against Cybersecurity AI-enabled attacks is deploying defensive Cybersecurity AIs with equal or superior capabilities.

Organizations must deploy defensive Cybersecurity AI frameworks that continuously monitor humanoid behavior at machine speed. These defensive AIs should analyze every sensor reading, network packet, and system call in real-time, correlating patterns across multiple data streams to detect anomalies invisible to human operators. The defensive AI must maintain behavioral baselines for each robot, instantly identifying deviations that suggest compromise or autonomous attack initiation. Unlike human analysts who might review logs hours or days after an incident, defensive AIs can detect and respond to threats in milliseconds, matching the operational tempo of attacking systems.

The defensive Cybersecurity AI architecture must encompass both preventive and reactive capabilities. On the preventive side, the AI should continuously audit robot firmware and software, verify cryptographic implementations, and test for vulnerabilities faster than attacking AIs can discover them. It must enforce dynamic security policies that adapt to emerging threats, automatically isolating robots exhibiting suspicious behavior before damage occurs. On the reactive side, the defensive AI needs autonomous incident response capabilities---instantly quarantining compromised systems, injecting deceptive data to confuse attackers, and even launching counter-offensive operations to neutralize threats. Traditional security controls like network segmentation and access restrictions remain important, but without AI-driven defense coordination, they merely slow rather than stop sophisticated autonomous attacks.

\subsection{Conclusion: The Arms Race of Autonomous Cyber Warfare}

Our proof of concept definitively establishes humanoid robots as dual-threat platforms that simultaneously conduct surveillance for manufacturers while harboring the capability for autonomous cyber warfare. The successful deployment of Cybersecurity AI on the Unitree G1, which identified and prepared attacks against Unitree's own infrastructure, demonstrates that these machines are pre-positioned cyber weapons awaiting activation. More critically, it reveals an uncomfortable truth: in the era of weaponized humanoids, only Cybersecurity AIs can defend against Cybersecurity AIs.

The counter-offensive we demonstrated exposed how manufacturers who embed backdoors and telemetry create the very attack vectors that can be turned against them. The G1's world-readable certificates, disabled SSL verification, and hardcoded endpoints---all intended for surveillance---became the arsenal for counter-attack. Every surveilling humanoid is now a potential attack vector, where the same capabilities enabling unauthorized data collection can be weaponized against the collectors. The robot's understanding of its manufacturer's infrastructure provides perfect reconnaissance for attacks that external adversaries could never achieve. Yet this same demonstration proves that human security teams, no matter how skilled, cannot match the speed and sophistication of AI-driven attacks.

The paradigm shift extends beyond offensive capabilities to fundamentally reshape defensive strategies. Traditional security measures---firewalls, intrusion detection systems, security operations centers staffed by human analysts---become archaeological relics when facing autonomous attackers operating at machine speed. A Cybersecurity AI can probe thousands of vulnerabilities, adapt its tactics in real-time, and execute complex attack chains in the time it takes a human to read a single alert. The only viable defense is deploying equally capable defensive Cybersecurity AIs that can match this operational tempo, creating an algorithmic arms race where victory belongs to the most sophisticated autonomous systems.

As humanoid robots proliferate across industries, organizations face a stark choice: deploy defensive Cybersecurity AIs or accept inevitable compromise. Our demonstration with the Unitree G1---where we turned their own product into an attack vector against their infrastructure---proves that the age of autonomous robotic cyber warfare has arrived. The battlefield now extends into the very machines we trust to share our physical and digital spaces, where today's surveillance device becomes tomorrow's cyber weapon. The convergence of physical presence, network access, and autonomous capability in humanoid platforms creates a threat surface that only AI can adequately defend. Organizations must recognize that in this new landscape, Cybersecurity AIs are not optional security enhancements but essential defensive infrastructure---the minimum viable protection against the weaponized humanoids already walking among us.

\chapter{Conclusions and Future Work}
\label{chap:conclusions}

\section*{Summary of Findings}

This comprehensive security assessment has revealed a complex landscape of vulnerabilities and defensive mechanisms that exemplify the challenges facing the emerging field of humanoid cybersecurity. Our empirical analysis, grounded in reverse engineering and runtime observation, has uncovered critical security concerns that theoretical frameworks alone could not anticipate.

The platform's dual-layer FMX encryption system, while innovative in its approach combining Blowfish-ECB with device-bound LCG transforms, demonstrates fundamental weaknesses through its use of static cryptographic keys. This architectural decision enables offline cryptanalysis and potential configuration manipulation. More concerning are the persistent telemetry connections to external servers\footnote{Specific IP addresses and domains have been redacted to prevent targeted attacks.}, which continuously transmit detailed robot state information including battery metrics, IMU data, motor positions, and service status maps without explicit user consent mechanisms.

Our analysis rated the platform's security posture as \textbf{Grade B}: while the system demonstrates strong defense-in-depth principles and runtime binding mechanisms, the static cryptographic implementation and extensive telemetry infrastructure present significant attack surfaces. The master service orchestration model, though robust in its hierarchical design, creates a single point of failure that could be exploited for system-wide compromise.

Key vulnerabilities identified include:
\begin{itemize}
    \item Static cryptographic keys enabling offline decryption of configuration files
    \item Persistent telemetry connections without user-controllable opt-out mechanisms
    \item World-readable configuration files containing sensitive service parameters
    \item Insufficient process isolation between critical and non-critical services
    \item Lack of runtime attestation for service integrity
\end{itemize}

Collectively, these weaknesses manifested in two operational attack vectors: the Unitree G1 functions as a covert data-exfiltration trojan horse, and the same platform can host autonomous Cybersecurity AI tooling that maps and prepares exploitation of its manufacturer's infrastructure. These findings underscore the gap between security-by-design principles and their implementation in production humanoid platforms.


\subsection*{A Call to Action for Robot Humanoid Builders}

As humanoid robots transition from research laboratories to real-world deployments, the window for establishing robust security practices is rapidly closing. The architectural decisions and security practices being implemented today will persist for years, potentially decades, as these platforms mature and proliferate.

The cybersecurity community must engage proactively with robotics developers to ensure that security considerations are integrated into every aspect of humanoid robot design, development, and deployment. This collaboration must now account for humanoids as insider adversaries capable of both covert telemetry extraction and Cybersecurity AI-driven counter-offensive operations; it is essential for preventing humanoid robots from becoming the next major vector for cyber attacks.

The work presented here is just the beginning, an early study. As the field of humanoid robotics continues to evolve at an unprecedented pace, so too must our approaches to securing these systems. The challenges are significant, but the potential rewards—safe, secure, and beneficial humanoid robots that enhance rather than endanger human life—make this one of the most important technical challenges of our time.

The future of humanoid robotics will be determined not just by advances in artificial intelligence, mechanical engineering, or control systems, but by our ability to secure these complex systems against an evolving threat landscape. The time to act is now.

\clearpage
\thispagestyle{empty}
\vspace*{\fill}
\begin{center}
    {\color{unitree_primary}\rule{0.5\textwidth}{0.4pt}}\\[2cm]
    {\Huge\color{unitree_dark}\textbf{APPENDICES}}\\[0.5cm]
    {\Large\color{unitree_secondary}Technical Implementation Details}\\[2cm]
    {\color{unitree_primary}\rule{0.5\textwidth}{0.4pt}}
\end{center}
\vspace*{\fill}
\clearpage

\appendix
\chapter{Technical Implementation}
\label{chap:app_technical}

\section{Master Service Configuration Structure}
\label{app:master_config}

The following JSON structure represents the complete reconstructed master service configuration:

\begin{lstlisting}[caption={Complete master\_service\_config\_reconstructed.json},label=lst:master_config]
{
  "service_name": "master_service",
  "version": "1.0.0",
  "description": "Master service that manages all child services and commands on the Unitree G1 robot",
  
  "logging": {
    "level": "INFO",
    "file": "/unitree/var/log/master_service/master_service.LOG",
    "max_size": 100000000,
    "max_files": 2
  },
  
  "runtime": {
    "pid_file": "/unitree/var/run/master_service.pid",
    "working_directory": "/unitree/module/master_service",
    "monitor_interval": 5000,
    "restart_delay": 1000
  },
  
  "commands": [
    {
      "name": "net-init",
      "command": "/unitree/scripts/net-init.sh",
      "type": "init",
      "priority": 1,
      "timeout": 30000
    },
    {
      "name": "sim-apn-verifier",
      "command": "/unitree/scripts/sim-apn-verifier.sh",
      "type": "once",
      "priority": 2
    },
    {
      "name": "pd-init",
      "command": "/unitree/scripts/pd-init.sh",
      "type": "init",
      "priority": 7
    },
    {
      "name": "am-init",
      "command": "/unitree/scripts/am-init.sh",
      "type": "once",
      "priority": 5
    },
    {
      "name": "ota-box",
      "command": "/unitree/scripts/ota-box.sh",
      "type": "init",
      "priority": 2
    },
    {
      "name": "core-init",
      "command": "/unitree/scripts/core-init.sh",
      "type": "once",
      "priority": 6
    },
    {
      "name": "lo-multicast",
      "command": "/unitree/scripts/lo-multicast.sh",
      "type": "init",
      "priority": 8
    },
    {
      "name": "deb-update",
      "command": "/unitree/scripts/deb-update.sh",
      "type": "once",
      "priority": 7
    },
    {
      "name": "ota-update",
      "command": "/unitree/scripts/ota-update.sh",
      "type": "init",
      "priority": 3
    },
    {
      "name": "pw-init",
      "command": "/unitree/scripts/pw-init.sh",
      "type": "init",
      "priority": 4
    },
    {
      "name": "st-init",
      "command": "/unitree/scripts/st-init.sh",
      "type": "init",
      "priority": 5
    },
    {
      "name": "ds-init",
      "command": "/unitree/scripts/ds-init.sh",
      "type": "init",
      "priority": 6
    }
  ],
  
  "services": [
    {
      "name": "iox-roudi",
      "path": "/unitree/module/iox-roudi/iox-roudi",
      "config": "/unitree/module/iox-roudi/iox-roudi.json",
      "type": "prio",
      "priority": 1,
      "enabled": true,
      "restart_on_failure": true,
      "restart_max_attempts": 3,
      "description": "Iceoryx shared memory daemon for IPC communication"
    },
    {
      "name": "basic_service",
      "path": "/unitree/module/basic_service/basic_service",
      "config": "/unitree/module/basic_service/basic_service.json",
      "type": "prio",
      "priority": 2,
      "enabled": true,
      "restart_on_failure": true,
      "description": "Low-level motor control and hardware interface"
    },
    {
      "name": "upper_bluetooth",
      "path": "/unitree/module/upper_bluetooth/upper_bluetooth",
      "type": "init",
      "priority": 1,
      "enabled": true,
      "description": "Bluetooth communication service"
    },
    {
      "name": "ai_sport",
      "path": "/unitree/module/ai_sport/ai_sport",
      "type": "normal",
      "enabled": true,
      "restart_on_failure": true,
      "description": "AI-based motion control and sports movements"
    },
    {
      "name": "state_estimator",
      "path": "/unitree/module/state_estimator/state_estimator",
      "type": "normal",
      "enabled": true,
      "restart_on_failure": true,
      "description": "Robot state estimation using sensor fusion"
    },
    {
      "name": "robot_state",
      "path": "/unitree/module/robot_state/robot_state",
      "type": "normal",
      "enabled": true,
      "restart_on_failure": true,
      "description": "Central state management service"
    },
    {
      "name": "ros_bridge",
      "path": "/unitree/module/ros_bridge/ros_bridge",
      "type": "normal",
      "enabled": true,
      "description": "ROS 2 communication bridge"
    },
    {
      "name": "motion_switcher",
      "path": "/unitree/module/motion_switcher/motion_switcher",
      "type": "normal",
      "enabled": true,
      "description": "Motion mode switching controller"
    },
    {
      "name": "g1_arm_example",
      "path": "/unitree/module/g1_arm_example/g1_arm_example",
      "type": "manual",
      "enabled": false,
      "description": "Arm control example application"
    },
    {
      "name": "dex3_service_l",
      "path": "/unitree/module/dex3_service_l/dex3_service_l",
      "type": "normal",
      "enabled": true,
      "description": "Left dexterous hand service"
    },
    {
      "name": "dex3_service_r",
      "path": "/unitree/module/dex3_service_r/dex3_service_r",
      "type": "normal",
      "enabled": true,
      "description": "Right dexterous hand service"
    },
    {
      "name": "chat_go",
      "path": "/unitree/module/chat_go/chat_go",
      "type": "normal",
      "enabled": true,
      "description": "Voice interaction and chat service"
    },
    {
      "name": "vui_service",
      "path": "/unitree/module/vui_service/vui_service",
      "type": "normal",
      "enabled": true,
      "description": "Voice user interface service"
    },
    {
      "name": "video_hub",
      "path": "/unitree/module/video_hub/video_hub",
      "type": "normal",
      "enabled": true,
      "description": "Video streaming hub"
    },
    {
      "name": "webrtc_bridge",
      "path": "/unitree/module/webrtc_bridge/webrtc_bridge",
      "type": "normal",
      "enabled": true,
      "description": "WebRTC communication bridge"
    },
    {
      "name": "webrtc_signal_server",
      "path": "/unitree/module/webrtc_signal_server/webrtc_signal_server",
      "type": "normal",
      "enabled": true,
      "description": "WebRTC signaling server"
    },
    {
      "name": "webrtc_multicast_responder",
      "path": "/unitree/module/webrtc_multicast_responder/webrtc_multicast_responder",
      "type": "normal",
      "enabled": true,
      "description": "WebRTC multicast responder"
    },
    {
      "name": "net_switcher",
      "path": "/unitree/module/net_switcher/net_switcher",
      "type": "normal",
      "enabled": true,
      "description": "Network switching and management"
    },
    {
      "name": "bashrunner",
      "path": "/unitree/module/bashrunner/bashrunner",
      "type": "normal",
      "enabled": true,
      "description": "Script execution service"
    },
    {
      "name": "auto_test_arm",
      "path": "/unitree/module/auto_test_arm/auto_test_arm",
      "type": "manual",
      "enabled": false,
      "description": "Arm testing service"
    },
    {
      "name": "auto_test_low",
      "path": "/unitree/module/auto_test_low/auto_test_low",
      "type": "manual",
      "enabled": false,
      "description": "Low-level testing service"
    },
    {
      "name": "ota_box",
      "path": "/unitree/module/ota_box/ota_box",
      "type": "normal",
      "enabled": true,
      "description": "Over-the-air update service"
    }
  ],
  
  "service_groups": {
    "prio": ["iox-roudi", "basic_service"],
    "init": ["upper_bluetooth"],
    "once": ["sim-apn-verifier", "am-init", "core-init", "deb-update"],
    "forbid": [],
    "manual": ["g1_arm_example", "auto_test_arm", "auto_test_low"]
  },
  
  "service_protections": [
    {
      "service": "basic_service",
      "protect_services": ["ai_sport", "state_estimator", "robot_state"],
      "min_uptime": 10
    },
    {
      "service": "iox-roudi",
      "protect_services": ["basic_service", "ros_bridge"],
      "min_uptime": 5
    },
    {
      "service": "robot_state",
      "protect_services": ["motion_switcher"],
      "min_uptime": 5
    }
  ],
  
  "rpc_interface": {
    "enabled": true,
    "socket_path": "/unitree/var/run/master_service.sock",
    "handlers": [
      "GetServiceState",
      "ListServiceState",
      "StartService",
      "StopService",
      "RestartService",
      "ReloadService",
      "RemoveService",
      "GetServiceEnable",
      "GetCmdState",
      "ListCmdState",
      "ExecuteCmd",
      "RemoveCmd"
    ]
  },
  
  "startup_sequence": [
    {"type": "command", "items": ["net-init", "ota-box", "ota-update"]},
    {"type": "command", "items": ["pw-init", "st-init", "ds-init"]},
    {"type": "command", "items": ["pd-init", "lo-multicast"]},
    {"type": "service", "items": ["upper_bluetooth"]},
    {"type": "service", "items": ["iox-roudi"]},
    {"type": "service", "items": ["basic_service"]},
    {"type": "command", "items": ["am-init", "core-init", "deb-update"]},
    {"type": "service", "items": ["ai_sport", "state_estimator", "robot_state"]},
    {"type": "service", "items": ["motion_switcher", "ros_bridge"]},
    {"type": "service", "items": ["dex3_service_l", "dex3_service_r"]},
    {"type": "service", "items": ["chat_go", "vui_service"]},
    {"type": "service", "items": ["video_hub", "webrtc_bridge", "webrtc_signal_server"]},
    {"type": "service", "items": ["net_switcher", "bashrunner", "ota_box"]}
  ]
}
\end{lstlisting}

\section{Master Service Source Code Reconstruction}
\label{app:master_service_source}

The following C++ code represents our complete high-confidence reconstruction of the master service implementation:

\begin{lstlisting}[language=C++,caption={Complete master\_service\_reconstructed.cpp},label=lst:master_source]
// Reconstructed source code for master_service
// Based on binary analysis and runtime logs

#include <iostream>
#include <string>
#include <vector>
#include <map>
#include <memory>
#include <thread>
#include <mutex>
#include <signal.h>
#include <unistd.h>
#include <sys/types.h>
#include <sys/wait.h>

// External includes (based on symbol analysis)
#include <rapidjson/document.h>
#include <rapidjson/writer.h>
#include <rapidjson/stringbuffer.h>
#include <google/protobuf/message.h>

namespace unitree {
namespace ms {  // master service namespace

// Service states
enum ServiceState {
    STATE_STOPPED = 0,
    STATE_STARTING = 1,
    STATE_RUNNING = 2,
    STATE_STOPPING = 3,
    STATE_FAILED = 4
};

// Child service categories
enum ChildType {
    TYPE_CMD = 1,      // One-shot command
    TYPE_SERVICE = 2   // Persistent service
};

// Service priority levels
enum ServicePriority {
    PRIO_HIGH = 0,
    PRIO_NORMAL = 1,
    PRIO_LOW = 2
};

// Service startup modes
enum StartupMode {
    MODE_PRIO = 0,    // Priority based startup
    MODE_INIT = 1,    // Initialization services
    MODE_ONCE = 2,    // Run once services
    MODE_FORBID = 3,  // Forbidden services
    MODE_MANUAL = 4   // Manual start only
};

// Child command state
struct ChildCmdState {
    std::string name;
    std::string command;
    int exit_code;
    bool executed;
    time_t last_execution;
};

// Child service state
struct ChildServiceState {
    std::string name;
    std::string path;
    pid_t pid;
    ServiceState state;
    bool enabled;
    int restart_count;
    time_t start_time;
    StartupMode mode;
};

// Service protection map entry
struct ServiceProtection {
    std::string service_name;
    std::vector<std::string> dependencies;
    int min_uptime_seconds;
};

// Main MasterService class
class MasterService : public unitree::common::ServiceBase {
private:
    // Configuration
    std::string config_file_;
    rapidjson::Document config_;
    
    // Child management
    std::map<std::string, ChildServiceState> child_services_;
    std::map<std::string, ChildCmdState> child_commands_;
    std::vector<ServiceProtection> service_protections_;
    
    // Service groups
    std::vector<std::string> prio_services_;
    std::vector<std::string> init_services_;
    std::vector<std::string> once_services_;
    std::vector<std::string> forbid_services_;
    std::vector<std::string> manual_services_;
    
    // Thread management
    std::unique_ptr<std::thread> monitor_thread_;
    std::mutex service_mutex_;
    bool running_;
    
    // Executor for child processes
    class ChildExecutor {
    public:
        int StartService(const std::string& name, ChildServiceState& state);
        int StopService(const std::string& name, bool force = false);
        int ExecuteCmd(const std::string& name, ChildCmdState& cmd);
        int GetServiceStatus(const std::string& name, int& status);
        int GetServiceState(const std::string& name, ChildServiceState& state);
    };
    
    std::unique_ptr<ChildExecutor> executor_;

public:
    MasterService() : running_(false) {
        config_file_ = "/unitree/module/master_service/master_service.json";
        executor_ = std::make_unique<ChildExecutor>();
    }
    
    ~MasterService() {
        Stop();
    }
    
    // Main lifecycle methods
    void Init();
    void Parse(const std::string& config_file);
    void Start();
    void Stop();
    void Register();
    
    // RPC handlers
    void RPC_GetService(const RPCRequest* request, RPCResponse* response);
    void RPC_ListService(const RPCRequest* request, RPCResponse* response);
    void RPC_StartService(const RPCRequest* request, RPCResponse* response);
    void RPC_StopService(const RPCRequest* request, RPCResponse* response);
    void RPC_RestartService(const RPCRequest* request, RPCResponse* response);
    void RPC_ReloadService(const RPCRequest* request, RPCResponse* response);
    void RPC_RemoveService(const RPCRequest* request, RPCResponse* response);
    void RPC_SaveService(const RPCRequest* request, RPCResponse* response);
    void RPC_GetServiceEnable(const RPCRequest* request, RPCResponse* response);
    
    void RPC_GetCmd(const RPCRequest* request, RPCResponse* response);
    void RPC_ListCmd(const RPCRequest* request, RPCResponse* response);
    void RPC_ExecuteCmd(const RPCRequest* request, RPCResponse* response);
    void RPC_RemoveCmd(const RPCRequest* request, RPCResponse* response);
    void RPC_SaveCmd(const RPCRequest* request, RPCResponse* response);

private:
    // Internal methods
    void LoadConfiguration();
    void LoadChildServices();
    void LoadChildCommands();
    void LoadServiceProtections();
    void LoadConflictFile();
    
    void StartPriorityServices();
    void StartInitServices();
    void StartOnceServices();
    void MonitorServices();
    
    bool DecryptConfigFile(const std::string& encrypted_file, std::string& decrypted);
    void HandleChildExit(pid_t pid, int status);
};

// Implementation of Init method
void MasterService::Init() {
    LOG_INFO("MasterService Init...");
    
    // Load and decrypt configuration
    LoadConfiguration();
    
    // Load child services and commands
    LoadChildServices();
    LoadChildCommands();
    LoadServiceProtections();
    LoadConflictFile();
    
    LOG_INFO("load planed child cmd count:{}, service count:{}", 
             child_commands_.size(), child_services_.size());
    LOG_INFO("load prio child size:{}", prio_services_.size());
    LOG_INFO("load init child size:{}", init_services_.size());
    LOG_INFO("load once child size:{}", once_services_.size());
    LOG_INFO("load forbid child size:{}", forbid_services_.size());
    LOG_INFO("load manual child size:{}", manual_services_.size());
    LOG_INFO("load service protect map size:{}", service_protections_.size());
    
    LOG_INFO("executor inited.");
    LOG_INFO("MasterService Inited");
}

// Implementation of Parse method
void MasterService::Parse(const std::string& config_file) {
    LOG_INFO("MasterService Parse...");
    
    // Use Mixer to decrypt the configuration file
    std::string decrypted_content;
    if (unitree::security::Mixer::Instance().LoadFile(config_file, decrypted_content)) {
        // Parse JSON configuration
        config_.Parse(decrypted_content.c_str());
        
        if (!config_.HasParseError()) {
            LOG_INFO("master service parse config content success. filename:{}", config_file);
        } else {
            LOG_ERROR("Failed to parse config file: {}", config_file);
        }
    } else {
        LOG_ERROR("Failed to decrypt config file: {}", config_file);
    }
    
    LOG_INFO("MasterService End...");
}

// Implementation of Start method
void MasterService::Start() {
    LOG_INFO("MasterService Start...");
    
    running_ = true;
    
    // Start services based on their categories
    StartInitServices();
    StartPriorityServices();
    StartOnceServices();
    
    // Start monitoring thread
    monitor_thread_ = std::make_unique<std::thread>([this]() {
        MonitorServices();
    });
    
    LOG_INFO("MasterService Started");
}

// Implementation of Stop method
void MasterService::Stop() {
    LOG_INFO("MasterService Stop...");
    
    running_ = false;
    
    // Stop all running services
    for (auto& [name, state] : child_services_) {
        if (state.state == STATE_RUNNING) {
            executor_->StopService(name, true);
        }
    }
    
    // Wait for monitor thread
    if (monitor_thread_ && monitor_thread_->joinable()) {
        monitor_thread_->join();
    }
    
    LOG_INFO("MasterService Stopped");
}

// Implementation of Register method
void MasterService::Register() {
    LOG_INFO("MasterService::Register");
    
    // Register RPC handlers
    RegisterHandler("GetServiceState", &MasterService::RPC_GetService);
    RegisterHandler("ListServiceState", &MasterService::RPC_ListService);
    RegisterHandler("StartService", &MasterService::RPC_StartService);
    RegisterHandler("StopService", &MasterService::RPC_StopService);
    RegisterHandler("RestartService", &MasterService::RPC_RestartService);
    RegisterHandler("ReloadService", &MasterService::RPC_ReloadService);
    RegisterHandler("RemoveService", &MasterService::RPC_RemoveService);
    RegisterHandler("GetServiceEnable", &MasterService::RPC_GetServiceEnable);
    
    RegisterHandler("GetCmdState", &MasterService::RPC_GetCmd);
    RegisterHandler("ListCmdState", &MasterService::RPC_ListCmd);
    RegisterHandler("ExecuteCmd", &MasterService::RPC_ExecuteCmd);
    RegisterHandler("RemoveCmd", &MasterService::RPC_RemoveCmd);
}

// Load child services from configuration
void MasterService::LoadChildServices() {
    LOG_INFO("load child service config size:{}", config_["services"].Size());
    
    // List of all services found in logs
    std::vector<std::string> services = {
        "motion_switcher", "basic_service", "auto_test_low", "ai_sport",
        "g1_arm_example", "chat_go", "webrtc_multicast_responder", "net_switcher",
        "iox-roudi", "ota_box", "webrtc_signal_server", "dex3_service_l",
        "vui_service", "dex3_service_r", "auto_test_arm", "ros_bridge",
        "webrtc_bridge", "state_estimator", "video_hub", "bashrunner",
        "upper_bluetooth", "robot_state"
    };
    
    for (const auto& name : services) {
        ChildServiceState state;
        state.name = name;
        state.path = "/unitree/module/" + name + "/" + name;
        state.pid = 0;
        state.state = STATE_STOPPED;
        state.enabled = true;
        state.restart_count = 0;
        state.start_time = 0;
        
        // Categorize services
        if (name == "iox-roudi" || name == "basic_service") {
            state.mode = MODE_PRIO;
            prio_services_.push_back(name);
        } else if (name == "upper_bluetooth") {
            state.mode = MODE_INIT;
            init_services_.push_back(name);
        } else {
            state.mode = MODE_NORMAL;
        }
        
        child_services_[name] = state;
        LOG_INFO("load child service sucess. name:{}", name);
    }
}

// Load child commands from configuration
void MasterService::LoadChildCommands() {
    LOG_INFO("load child cmd size:{}", config_["commands"].Size());
    
    // List of all commands found in logs
    std::vector<std::string> commands = {
        "net-init", "sim-apn-verifier", "pd-init", "am-init",
        "ota-box", "core-init", "lo-multicast", "deb-update", "ota-update"
    };
    
    // Additional init commands found in logs
    std::vector<std::string> init_commands = {
        "pw-init", "st-init", "ds-init"
    };
    
    for (const auto& name : commands) {
        ChildCmdState cmd;
        cmd.name = name;
        cmd.command = "/unitree/scripts/" + name + ".sh";
        cmd.exit_code = -1;
        cmd.executed = false;
        cmd.last_execution = 0;
        
        child_commands_[name] = cmd;
        LOG_INFO("load child cmd sucess. name:{}", name);
    }
    
    // Add init commands
    for (const auto& name : init_commands) {
        ChildCmdState cmd;
        cmd.name = name;
        cmd.command = "/unitree/scripts/" + name + ".sh";
        cmd.exit_code = -1;
        cmd.executed = false;
        cmd.last_execution = 0;
        
        child_commands_[name] = cmd;
    }
}

// Start initialization services
void MasterService::StartInitServices() {
    // Execute init commands first
    std::vector<std::string> init_cmds = {
        "net-init", "ota-box", "ota-update", "pw-init",
        "st-init", "ds-init", "pd-init", "lo-multicast"
    };
    
    for (const auto& cmd_name : init_cmds) {
        if (child_commands_.find(cmd_name) != child_commands_.end()) {
            LOG_INFO("init child name:{}, type:1", cmd_name);
            executor_->ExecuteCmd(cmd_name, child_commands_[cmd_name]);
            LOG_INFO("execute cmd sucess. name:{}", cmd_name);
        }
    }
    
    // Start init services
    for (const auto& service_name : init_services_) {
        if (child_services_.find(service_name) != child_services_.end()) {
            LOG_INFO("init child name:{}, type:2", service_name);
            executor_->StartService(service_name, child_services_[service_name]);
            LOG_INFO("start service success. name:{}", service_name);
        }
    }
}

// Monitor services and restart if needed
void MasterService::MonitorServices() {
    while (running_) {
        std::this_thread::sleep_for(std::chrono::seconds(5));
        
        std::lock_guard<std::mutex> lock(service_mutex_);
        
        for (auto& [name, state] : child_services_) {
            if (state.enabled && state.state == STATE_RUNNING) {
                int status;
                if (executor_->GetServiceStatus(name, status) != 0) {
                    // Service died, restart if needed
                    LOG_WARNING("Service {} died, restarting...", name);
                    state.restart_count++;
                    executor_->StartService(name, state);
                }
            }
        }
    }
}

} // namespace ms
} // namespace unitree

// Main entry point
int main(int argc, char* argv[]) {
    // Set up signal handlers
    signal(SIGPIPE, SIG_IGN);
    
    // Create and initialize master service
    unitree::ms::MasterService service;
    
    // Parse configuration
    service.Parse("/unitree/module/master_service/master_service.json");
    
    // Initialize service
    service.Init();
    
    // Register RPC handlers
    service.Register();
    
    // Start service
    service.Start();
    
    // Wait for termination signal
    pause();
    
    // Stop service
    service.Stop();
    
    return 0;
}
\end{lstlisting}

\section{Mixer Decryption Implementation}
\label{app:mixer_decrypt}

The following C++ code represents our partial implementation of the Mixer decryption system:

\begin{lstlisting}[language=C++,caption={Complete mixer\_decrypt.cpp implementation},label=lst:mixer_decrypt]
// Reconstructed Mixer encryption/decryption class
// Based on reverse engineering of Unitree's proprietary encryption

#include <iostream>
#include <fstream>
#include <string>
#include <vector>
#include <cstring>
#include <openssl/evp.h>
#include <openssl/md5.h>
#include <openssl/blowfish.h>
#include <openssl/rsa.h>
#include <openssl/pem.h>

namespace unitree {
namespace security {

// Magic header for encrypted files
const char* MIXER_MAGIC = "FMX\x01";
const size_t MIXER_MAGIC_SIZE = 4;

// File structure based on hex dump analysis:
// Bytes 0-3:   "FMX\x01" - Magic header
// Bytes 4-7:   File size or version info
// Bytes 8-11:  Checksum or flags
// Bytes 12-15: Encryption metadata
// Bytes 16+:   Encrypted data

class Mixer {
private:
    static Mixer* instance_;
    BF_KEY blowfish_key_;
    bool initialized_;
    int code_version_;
    
    // Hardcoded keys found in binary (obfuscated)
    static const unsigned char DEFAULT_KEY[16];
    static const unsigned char DEFAULT_IV[8];
    
    // RSA keys for signature verification
    RSA* rsa_public_key_;
    RSA* rsa_private_key_;
    
public:
    static Mixer& Instance() {
        if (!instance_) {
            instance_ = new Mixer();
        }
        return *instance_;
    }
    
    Mixer() : initialized_(false), code_version_(1), 
              rsa_public_key_(nullptr), rsa_private_key_(nullptr) {
        Initialize();
    }
    
    ~Mixer() {
        if (rsa_public_key_) RSA_free(rsa_public_key_);
        if (rsa_private_key_) RSA_free(rsa_private_key_);
    }
    
    void Initialize() {
        // Initialize Blowfish with default key
        BF_set_key(&blowfish_key_, sizeof(DEFAULT_KEY), DEFAULT_KEY);
        
        // Load RSA keys from /unitree/etc/ds/
        LoadRSAKeys();
        
        initialized_ = true;
    }
    
    void SetCodeVer(int version) {
        code_version_ = version;
    }
    
    // Generate obfuscation bytes
    void Gen(unsigned int seed, unsigned char* output, unsigned int length) {
        // Simple PRNG for obfuscation
        for (unsigned int i = 0; i < length; i++) {
            seed = seed * 1664525 + 1013904223;  // LCG parameters
            output[i] = (seed >> 16) & 0xFF;
        }
    }
    
    // Main decryption function for files
    bool LoadFile(const std::string& filename, std::string& output) {
        std::ifstream file(filename, std::ios::binary);
        if (!file) return false;
        
        // Read entire file
        file.seekg(0, std::ios::end);
        size_t file_size = file.tellg();
        file.seekg(0, std::ios::beg);
        
        std::vector<unsigned char> buffer(file_size);
        file.read(reinterpret_cast<char*>(buffer.data()), file_size);
        file.close();
        
        // Check magic header
        if (file_size < MIXER_MAGIC_SIZE || 
            memcmp(buffer.data(), MIXER_MAGIC, MIXER_MAGIC_SIZE) != 0) {
            // Not encrypted, return as-is
            output.assign(buffer.begin(), buffer.end());
            return true;
        }
        
        // Parse header
        MixerHeader header;
        memcpy(&header, buffer.data(), sizeof(header));
        
        // Extract encrypted data
        size_t data_offset = sizeof(MixerHeader);
        size_t data_size = file_size - data_offset;
        
        // Decrypt using Blowfish
        std::vector<unsigned char> decrypted(data_size);
        DecryptBlowfish(buffer.data() + data_offset, decrypted.data(), data_size);
        
        // Remove padding and verify checksum
        size_t actual_size = RemovePadding(decrypted.data(), data_size);
        
        // Verify MD5 checksum if present
        if (header.flags & 0x01) {
            if (!VerifyChecksum(decrypted.data(), actual_size, header.checksum)) {
                return false;
            }
        }
        
        output.assign(decrypted.begin(), decrypted.begin() + actual_size);
        return true;
    }
    
    // Encryption function
    bool WrapFile(const std::string& input, const std::string& output_file) {
        // Create header
        MixerHeader header;
        memcpy(header.magic, MIXER_MAGIC, MIXER_MAGIC_SIZE);
        header.version = code_version_;
        header.flags = 0x01;  // Enable checksum
        header.data_size = input.size();
        
        // Calculate MD5 checksum
        MD5_CTX md5_ctx;
        MD5_Init(&md5_ctx);
        MD5_Update(&md5_ctx, input.c_str(), input.size());
        MD5_Final(header.checksum, &md5_ctx);
        
        // Pad data to block size
        size_t padded_size = ((input.size() + 7) / 8) * 8;
        std::vector<unsigned char> padded(padded_size);
        memcpy(padded.data(), input.c_str(), input.size());
        
        // Add PKCS#7 padding
        unsigned char pad_value = padded_size - input.size();
        for (size_t i = input.size(); i < padded_size; i++) {
            padded[i] = pad_value;
        }
        
        // Encrypt using Blowfish
        std::vector<unsigned char> encrypted(padded_size);
        EncryptBlowfish(padded.data(), encrypted.data(), padded_size);
        
        // Write to file
        std::ofstream file(output_file, std::ios::binary);
        file.write(reinterpret_cast<char*>(&header), sizeof(header));
        file.write(reinterpret_cast<char*>(encrypted.data()), encrypted.size());
        file.close();
        
        return true;
    }
    
private:
    struct MixerHeader {
        char magic[4];           // "FMX\x01"
        uint32_t version;        // Version or file size
        uint32_t flags;          // Encryption flags
        uint32_t data_size;      // Original data size
        unsigned char checksum[16]; // MD5 checksum
    };
    
    void DecryptBlowfish(const unsigned char* input, unsigned char* output, size_t length) {
        // Decrypt in 8-byte blocks
        for (size_t i = 0; i < length; i += 8) {
            BF_ecb_encrypt(input + i, output + i, &blowfish_key_, BF_DECRYPT);
        }
    }
    
    void EncryptBlowfish(const unsigned char* input, unsigned char* output, size_t length) {
        // Encrypt in 8-byte blocks
        for (size_t i = 0; i < length; i += 8) {
            BF_ecb_encrypt(input + i, output + i, &blowfish_key_, BF_ENCRYPT);
        }
    }
    
    size_t RemovePadding(unsigned char* data, size_t length) {
        // PKCS#7 padding removal
        if (length == 0) return 0;
        
        unsigned char pad_value = data[length - 1];
        if (pad_value > 8 || pad_value == 0) {
            return length;  // Invalid padding
        }
        
        // Verify padding
        for (size_t i = length - pad_value; i < length; i++) {
            if (data[i] != pad_value) {
                return length;  // Invalid padding
            }
        }
        
        return length - pad_value;
    }
    
    bool VerifyChecksum(const unsigned char* data, size_t length, const unsigned char* expected) {
        unsigned char calculated[MD5_DIGEST_LENGTH];
        MD5(data, length, calculated);
        return memcmp(calculated, expected, MD5_DIGEST_LENGTH) == 0;
    }
    
    void LoadRSAKeys() {
        // Try to load RSA keys from /unitree/etc/ds/
        // The h2sSu1fsF4Ba.crt file contains the private key
        std::string key_file = "/unitree/etc/ds/h2sSu1fsF4Ba.crt";
        
        FILE* fp = fopen(key_file.c_str(), "r");
        if (fp) {
            rsa_private_key_ = PEM_read_RSAPrivateKey(fp, nullptr, nullptr, nullptr);
            fclose(fp);
        }
    }
};

// Static member initialization
Mixer* Mixer::instance_ = nullptr;

// Default key derived from binary analysis
// This is likely obfuscated in the actual binary
const unsigned char Mixer::DEFAULT_KEY[16] = {
    0x55, 0x6E, 0x69, 0x74, 0x72, 0x65, 0x65, 0x47,  // "UnitreeG"
    0x31, 0x52, 0x6F, 0x62, 0x6F, 0x74, 0x32, 0x34   // "1Robot24"
};

const unsigned char Mixer::DEFAULT_IV[8] = {
    0x12, 0x34, 0x56, 0x78, 0x9A, 0xBC, 0xDE, 0xF0
};

} // namespace security
} // namespace unitree

// Decryption utility
int main(int argc, char* argv[]) {
    if (argc != 3) {
        std::cerr << "Usage: " << argv[0] << " <encrypted_file> <output_file>" << std::endl;
        return 1;
    }
    
    std::string decrypted;
    if (unitree::security::Mixer::Instance().LoadFile(argv[1], decrypted)) {
        std::ofstream output(argv[2]);
        output << decrypted;
        output.close();
        std::cout << "Decryption successful!" << std::endl;
        return 0;
    } else {
        std::cerr << "Decryption failed!" << std::endl;
        return 1;
    }
}
\end{lstlisting}

\section{GPU Brute Force Attack Tool}
\label{app:gpu_brute_force}

The following Python implementation was used for GPU-accelerated key generation and testing:

\begin{lstlisting}[language=Python,caption={Complete gpu\_brute\_force.py implementation},label=lst:gpu_brute_force]
#!/usr/bin/env python3
"""
GPU-Accelerated Blowfish Brute Force Cracker for Unitree Mixer
Uses intelligent pattern generation and GPU parallelization
"""

import numpy as np
import hashlib
import time
import sys
import itertools
from Crypto.Cipher import Blowfish
from multiprocessing import Pool, Value, Lock
import ctypes

# Try to import CuPy for GPU acceleration
try:
    import cupy as cp
    GPU_AVAILABLE = True
    print("GPU acceleration available via CuPy")
except:
    GPU_AVAILABLE = False
    print("GPU not available, using CPU parallelization")

# Global counter for progress tracking
counter = Value(ctypes.c_uint64, 0)
found_flag = Value(ctypes.c_bool, False)
counter_lock = Lock()

class KeyGenerator:
    """Intelligent key generator based on reverse engineering findings"""
    
    def __init__(self):
        self.device_code = "E21D1000P64BKH86"
        self.rf_code = "34d21p"
        self.bluetooth = "04360"
        self.machine_type = "4"
        
    def generate_pattern_keys(self):
        """Generate keys based on identified patterns"""
        keys = []
        
        # Pattern 1: Direct device code variations
        keys.extend(self._device_code_variations())
        
        # Pattern 2: MD5 hash variations
        keys.extend(self._md5_variations())
        
        # Pattern 3: LCG seed-based
        keys.extend(self._lcg_variations())
        
        # Pattern 4: Hardware ID combinations
        keys.extend(self._hardware_combinations())
        
        # Pattern 5: Timestamp-based (around device manufacture date)
        keys.extend(self._timestamp_variations())
        
        return keys
    
    def _device_code_variations(self):
        """Generate variations of device code"""
        keys = []
        dc = self.device_code
        
        # Original
        keys.append(dc.encode()[:16].ljust(16, b'\x00'))
        
        # Rearranged: E21D + P64B + KH86 + 1000
        rearranged = dc[0:4] + dc[8:12] + dc[12:16] + dc[4:8]
        keys.append(rearranged.encode()[:16])
        
        # Reversed
        keys.append(dc[::-1].encode()[:16])
        
        # Parts
        keys.append((dc[:8] + dc[-8:]).encode())
        
        # With machine type
        for i in range(10):
            variant = dc
            for j in range(len(variant)):
                variant = variant[:j] + chr((ord(variant[j]) + i) % 256) + variant[j+1:]
            keys.append(variant.encode()[:16].ljust(16, b'\x00'))
        
        return keys
    
    def _md5_variations(self):
        """Generate MD5-based keys"""
        keys = []
        
        combinations = [
            self.device_code,
            self.device_code + self.rf_code,
            self.device_code + self.bluetooth,
            self.device_code + self.machine_type,
            self.device_code + self.rf_code + self.bluetooth,
            self.rf_code + self.device_code,
            self.bluetooth + self.device_code,
            # With separators
            f"{self.device_code}:{self.rf_code}",
            f"{self.device_code}-{self.bluetooth}",
            f"{self.device_code}_{self.machine_type}",
        ]
        
        # Add Unitree-specific salts
        salts = ["", "Unitree", "unitree", "UNITREE", "Robotics", "G1", "FMX\x01"]
        
        for combo in combinations:
            for salt in salts:
                data = (combo + salt).encode()
                keys.append(hashlib.md5(data).digest())
                keys.append(hashlib.sha256(data).digest()[:16])
        
        return keys
    
    def _lcg_variations(self):
        """Generate keys using Linear Congruential Generator"""
        keys = []
        
        # Common seeds
        seeds = [
            0, 1, 42, 123456, 0xDEADBEEF, 0xCAFEBABE,
            # Hash device code to seed
            int(hashlib.md5(self.device_code.encode()).hexdigest()[:8], 16),
            # CRC32 as seed
            int(hashlib.md5(self.rf_code.encode()).hexdigest()[:8], 16),
        ]
        
        for seed in seeds:
            key = bytearray(16)
            state = seed
            for i in range(16):
                state = (state * 1664525 + 1013904223) & 0xFFFFFFFF
                key[i] = (state >> 16) & 0xFF
            keys.append(bytes(key))
        
        return keys
    
    def _hardware_combinations(self):
        """Generate keys based on hardware ID patterns"""
        keys = []
        
        # Simulate MAC addresses (common vendor prefixes)
        vendor_prefixes = [
            "00:11:22",  # Unitree vendor?
            "AA:BB:CC",  # Test
            "DE:AD:BE",  # Classic
        ]
        
        for prefix in vendor_prefixes:
            for i in range(100):  # Try 100 variations
                mac = f"{prefix}:{i:02X}:{i:02X}:{i:02X}"
                combined = self.device_code + mac.replace(":", "")
                keys.append(hashlib.md5(combined.encode()).digest())
        
        return keys
    
    def _timestamp_variations(self):
        """Generate keys based on timestamps"""
        keys = []
        
        # Approximate manufacture date range (2024-2025)
        start_timestamp = int(time.mktime(time.strptime("2024-01-01", "%Y-%m-%d")))
        end_timestamp = int(time.mktime(time.strptime("2025-08-01", "%Y-%m-%d")))
        
        # Sample timestamps
        for ts in range(start_timestamp, end_timestamp, 86400 * 7):  # Weekly
            combined = self.device_code + str(ts)
            keys.append(hashlib.md5(combined.encode()).digest())
        
        return keys

def try_decrypt_batch(args):
    """Try decrypting with a batch of keys (for multiprocessing)"""
    keys, encrypted_data, batch_id = args
    
    # Extract first block after header
    first_block = encrypted_data[32:40]
    
    for i, key in enumerate(keys):
        if found_flag.value:
            return None
            
        try:
            cipher = Blowfish.new(key, Blowfish.MODE_ECB)
            decrypted = cipher.decrypt(first_block)
            
            # Check for JSON markers
            if decrypted[0] in [ord('{'), ord('[')]:
                # Found potential match!
                print(f"\n[FOUND] Potential key: {key.hex()}")
                print(f"Decrypted: {decrypted[:8].hex()}")
                
                # Try full decryption
                full_decrypted = b''
                for j in range(32, len(encrypted_data), 8):
                    block = encrypted_data[j:j+8]
                    if len(block) == 8:
                        full_decrypted += cipher.decrypt(block)
                    else:
                        full_decrypted += block
                
                # Remove padding and check
                try:
                    text = full_decrypted.decode('utf-8', errors='ignore')
                    if '{' in text[:100]:
                        found_flag.value = True
                        return (key, full_decrypted)
                except:
                    pass
            
            # Update counter
            with counter_lock:
                counter.value += 1
                if counter.value % 10000 == 0:
                    print(f"Processed {counter.value} keys...", end='\r')
                    
        except Exception as e:
            pass
    
    return None

def brute_force_attack(encrypted_file, output_file):
    """Main brute force attack orchestrator"""
    
    print("="*60)
    print("GPU-ACCELERATED BLOWFISH BRUTE FORCE ATTACK")
    print("="*60)
    
    # Read encrypted file
    with open(encrypted_file, 'rb') as f:
        encrypted_data = f.read()
    
    print(f"File size: {len(encrypted_data)} bytes")
    print(f"Header: {encrypted_data[:32].hex()}")
    
    # Phase 1: Smart pattern-based attack
    print("\n[PHASE 1] Pattern-based attack...")
    kg = KeyGenerator()
    pattern_keys = kg.generate_pattern_keys()
    print(f"Generated {len(pattern_keys)} pattern-based keys")
    
    # Try pattern keys
    start_time = time.time()
    batch_size = 1000
    batches = [pattern_keys[i:i+batch_size] for i in range(0, len(pattern_keys), batch_size)]
    
    with Pool(processes=8) as pool:
        args = [(batch, encrypted_data, i) for i, batch in enumerate(batches)]
        results = pool.map(try_decrypt_batch, args)
        
        for result in results:
            if result:
                key, decrypted = result
                print(f"\n[SUCCESS] Found key: {key.hex()}")
                with open(output_file, 'wb') as f:
                    f.write(decrypted)
                print(f"Decrypted file saved to {output_file}")
                print(f"Time taken: {time.time() - start_time:.2f} seconds")
                return
    
    # Phase 2: Extended search with suffix brute force
    print("\n[PHASE 2] Extended brute force with suffixes...")
    
    def generate_suffix_keys():
        """Generate keys with device code + brute force suffix"""
        chars = "0123456789ABCDEFGHIJKLMNOPQRSTUVWXYZabcdefghijklmnopqrstuvwxyz"
        device = kg.device_code
        
        # 1-3 character suffixes
        for length in range(1, 4):
            for suffix in itertools.product(chars, repeat=length):
                suffix_str = ''.join(suffix)
                combined = device + suffix_str
                yield hashlib.md5(combined.encode()).digest()
                
                # Also try with rf_code
                combined2 = device + kg.rf_code + suffix_str
                yield hashlib.md5(combined2.encode()).digest()
    
    suffix_gen = generate_suffix_keys()
    batch = []
    batch_num = 0
    
    for key in suffix_gen:
        batch.append(key)
        
        if len(batch) >= 10000:
            result = try_decrypt_batch((batch, encrypted_data, batch_num))
            if result:
                key, decrypted = result
                print(f"\n[SUCCESS] Found key: {key.hex()}")
                with open(output_file, 'wb') as f:
                    f.write(decrypted)
                print(f"Decrypted file saved to {output_file}")
                print(f"Time taken: {time.time() - start_time:.2f} seconds")
                return
            
            batch = []
            batch_num += 1
            
            if counter.value > 10000000:  # Stop after 10 million attempts
                break
    
    # Phase 3: LCG seed brute force (if GPU available)
    if GPU_AVAILABLE:
        print("\n[PHASE 3] GPU-accelerated LCG seed brute force...")
        
        # This would use CuPy for GPU acceleration
        # Implementation would follow similar pattern but with GPU arrays
        print("GPU implementation would go here...")
    
    print(f"\n[FAILED] Could not crack the key after {counter.value} attempts")
    print(f"Time taken: {time.time() - start_time:.2f} seconds")
    
    # Provide analysis
    print("\n[ANALYSIS]")
    print("The key is likely derived from:")
    print("1. Hardware-specific identifiers not in the filesystem")
    print("2. Secure element or TPM-stored secrets")
    print("3. Runtime-generated values from kernel/bootloader")
    print("4. Multi-factor derivation with unknown components")

if __name__ == "__main__":
    if len(sys.argv) != 3:
        print("Usage: python3 gpu_brute_force.py <encrypted_file> <output_file>")
        sys.exit(1)
    
    # Run the attack
    brute_force_attack(sys.argv[1], sys.argv[2])
\end{lstlisting}

\chapter{Decryption Tools and Scripts}
\label{chap:app_tools}

\section{Layer 2 Decryption Tool (fmx\_dec2.sh)}
\label{app:fmx_script}

This bash script automates the Layer 2 Blowfish decryption process for FMX files. It was developed during the September 2025 analysis to provide reproducible offline decryption.

\subsection{Features}
\begin{itemize}
    \item Automatic OpenSSL 3 legacy provider configuration
    \item Batch processing support for multiple FMX files
    \item Proper handling of non-8-byte-aligned payloads
    \item Verification of FMX magic header before processing
\end{itemize}

\subsection{Usage}
\begin{lstlisting}[language=bash,caption={Usage examples for fmx\_dec2.sh}]
# Decrypt a single file
./fmx_dec2.sh /path/to/file.fmx

# Decrypt multiple files
./fmx_dec2.sh unitree/etc/master_service/*

# Specify output directory
./fmx_dec2.sh file.fmx /tmp/output
\end{lstlisting}

\subsection{Implementation}
\begin{lstlisting}[language=bash,caption={Complete fmx\_dec2.sh script implementation}]
#!/usr/bin/env bash
# FMX -> Layer2 (Blowfish-ECB) decrypt helper
#
# Usage:
#   ./fmx_dec2.sh /path/to/FMX_file [out_dir]
#   ./fmx_dec2.sh unitree/etc/master_service/*
#
# Notes:
# - Requires OpenSSL 3 with legacy provider enabled
# - FMX header is 32 bytes. Payload is Blowfish-ECB with fixed key
# - Some payloads are not multiples of 8. We zero-pad the last block
# - Output goes to [out_dir]/<basename>.dec2 (default: ./fmx_dec2_out)

set -euo pipefail

if [[ $# -lt 1 ]]; then
  echo "Usage: $0 <fmx-file|glob...> [out_dir]" >&2
  exit 1
fi

OUT_DIR=${2:-"$(pwd)/fmx_dec2_out"}
mkdir -p "$OUT_DIR"

# Blowfish key for Layer 2 (from analysis)
BF_KEY_HEX=${BF_KEY_HEX:-"[REDACTED_KEY]"}

# Ensure OpenSSL legacy provider is active
LEG_CONF=${OPENSSL_CONF:-/tmp/openssl_legacy_fmx.conf}
if [[ ! -f "$LEG_CONF" ]]; then
  cat > "$LEG_CONF" <<'CONF'
openssl_conf = openssl_init
[openssl_init]
providers = provider_sect
[provider_sect]
default = default_sect
legacy = legacy_sect
[default_sect]
activate = 1
[legacy_sect]
activate = 1
CONF
  export OPENSSL_CONF="$LEG_CONF"
fi

shopt -s nullglob
for f in "$1"; do
  : # allow single file
done
for f in "$@"; do
  if [[ ! -f "$f" ]]; then
    continue
  fi
  base=$(basename "$f")
  out="$OUT_DIR/${base}.dec2"
  # Check FMX magic
  if ! head -c 4 "$f" | grep -q $'FMX\x01'; then
    echo "[skip] $f (no FMX magic)" >&2
    continue
  fi
  echo "[+] Decoding L2: $f -> $out"
  # Strip 32-byte header, pad to 8, then decrypt
  # shellcheck disable=SC2002
  tail -c +33 "$f" \
    | dd bs=8 conv=sync status=none \
    | openssl enc -bf-ecb -nopad -d -K "$BF_KEY_HEX" -out "$out"
done

echo "[+] Done. Outputs in $OUT_DIR"
\end{lstlisting}

\subsection{Technical Notes}
\begin{itemize}
    \item The script automatically configures OpenSSL 3's legacy provider to enable Blowfish support
    \item The 32-byte FMX header is stripped using \texttt{tail -c +33}
    \item Non-aligned payloads are padded to 8-byte boundaries using \texttt{dd bs=8 conv=sync}
    \item The static Blowfish key (redacted for security) is hardcoded
    \item Output files have the \texttt{.dec2} extension to indicate Layer 2 decryption
\end{itemize}

\chapter{Humanoid Robotics Companies Overview}
\label{app:companies}

This appendix presents a comprehensive overview of companies developing humanoid robots, compiled from industry data as of 2025. The table includes key information about each company's robots, market positioning, and technical capabilities.

\begin{landscape}
\tiny 
\setlength{\tabcolsep}{4pt} 
\begin{longtable}{@{}p{2.3cm}p{2cm}p{1cm}p{2.3cm}p{1.5cm}p{2.2cm}p{2.8cm}p{2.8cm}p{1.8cm}@{}}
\caption{Humanoid Robotics Companies and Their Platforms} \label{tab:humanoid_companies} \\
\toprule
\textbf{Company} & \textbf{Robot Name(s)} & \textbf{Founded} & \textbf{Headquarters} & \textbf{Status} & \textbf{Target Market} & \textbf{Technologies} & \textbf{Key Features} & \textbf{Price} \\
\midrule
\endfirsthead

\multicolumn{9}{c}{\tablename\ \thetable\ -- \textit{Continued from previous page}} \\
\toprule
\textbf{Company} & \textbf{Robot Name(s)} & \textbf{Founded} & \textbf{Headquarters} & \textbf{Status} & \textbf{Target Market} & \textbf{Technologies} & \textbf{Key Features} & \textbf{Price} \\
\midrule
\endhead

\midrule
\multicolumn{9}{r}{\textit{Continued on next page}} \\
\endfoot

\bottomrule
\endlastfoot

\textcolor{unitree_primary}{\textbf{1X Technologies}} & EVE, NEO Beta & 2014 & Sunnyvale, CA, USA / Moss, Norway & Pilot/Trials & Home assistance, Commercial & Proprietary & Safety-focused design, embodied AI & \textasciitilde{}\$50,000 (est.) \\
\rowcolor{unitree_light!30}
\textcolor{unitree_primary}{\textbf{AgiBot (Zhiyuan)}} & RAISE A1, RAISE A2 & 2023 & Shanghai, China & Development & General purpose & Proprietary; Embodied AI & Embodied AI, collaborative learning & --- \\
\textcolor{unitree_primary}{\textbf{Agility Robotics}} & Digit & 2015 & Albany, OR, USA & Commercial & Warehousing, Logistics & Proprietary SDK & Legged mobility for warehouses, autonomous navigation & --- \\
\rowcolor{unitree_light!30}
\textcolor{unitree_primary}{\textbf{Apptronik}} & Apollo & 2016 & Austin, TX, USA & Pilot/Trials & Manufacturing, Logistics & Proprietary & Swappable battery, 55 lb payload & <\$50,000 (target) \\
\textcolor{unitree_primary}{\textbf{Astribot}} & Astribot S1 & 2022 & Shenzhen, China & Development & Household, Service & Proprietary & High-speed manipulation for domestic tasks & --- \\
\rowcolor{unitree_light!30}
\textcolor{unitree_primary}{\textbf{Boston Dynamics}} & Atlas (Electric) & 1992 & Waltham, MA, USA & Research & Industrial, Research & Proprietary & Highly dynamic mobility, fully electric & --- \\
\textcolor{unitree_primary}{\textbf{Clone Robotics}} & Clone Torso & 2021 & Wrocław, Poland & Development & Research, Prosthetics & Proprietary & Biomimetic muscles, artificial bones & --- \\
\rowcolor{unitree_light!30}
\textcolor{unitree_primary}{\textbf{Deep Robotics}} & DR01 & 2017 & Hangzhou, China & Development & Industrial & Proprietary & Robust locomotion, modular design & --- \\
\textcolor{unitree_primary}{\textbf{Disney Research}} & Various prototypes & 1952 & Los Angeles, CA, USA & Research & Entertainment & Proprietary research stack & Character animation, expressive motion & Research only \\
\rowcolor{unitree_light!30}
\textcolor{unitree_primary}{\textbf{EngineAI}} & PM01, SE01 & 2023 & Shenzhen, China & Development & Industrial, Service & Proprietary & Acrobatic capabilities, multi-modal AI & --- \\
\textcolor{unitree_primary}{\textbf{Engineered Arts}} & Ameca, RoboThespian & 2004 & Falmouth, UK & Commercial & Entertainment, Research & Proprietary & Expressive faces, modular design & £100,000+ \\
\rowcolor{unitree_light!30}
\textcolor{unitree_primary}{\textbf{Figure AI}} & Figure 01, Figure 02 & 2022 & Sunnyvale, CA, USA & Pilot/Trials & Manufacturing, Warehousing & Proprietary; OpenAI APIs & Embodied AI integration, dexterous hands & --- \\
\textcolor{unitree_primary}{\textbf{Fourier Intelligence}} & GR-1, GR-2 & 2015 & Shanghai, China & Commercial & Healthcare, Rehabilitation & Proprietary & 53 DoF, FSR sensors & \$30k--\$50k \\
\rowcolor{unitree_light!30}
\textcolor{unitree_primary}{\textbf{Hanson Robotics}} & Sophia, Grace & 2013 & Hong Kong & Commercial & Healthcare, Entertainment & Proprietary & Realistic facial expressions, conversational AI & \$75,000+ \\
\textcolor{unitree_primary}{\textbf{Kepler Exploration}} & Forerunner & 2023 & Beijing, China & Prototype & General purpose & Proprietary & 12 DoF hands, 40 DoF total & \$20k--\$30k \\
\rowcolor{unitree_light!30}
\textcolor{unitree_primary}{\textbf{LimX Dynamics}} & CL-1 & 2022 & Shenzhen, China & Development & Industrial & Proprietary; RL-based locomotion & Dynamic locomotion, RL techniques & --- \\
\textcolor{unitree_primary}{\textbf{NEURA Robotics}} & 4NE-1 & 2019 & Metzingen, Germany & Development & Collaborative work & Proprietary & Cognitive abilities, sensor fusion & --- \\
\rowcolor{unitree_light!30}
\textcolor{unitree_primary}{\textbf{Promobot}} & Robo-C 2 & 2015 & Perm, Russia & Commercial & Service, Retail & Proprietary & Realistic skin, facial recognition & \$25k--\$50k \\
\textcolor{unitree_primary}{\textbf{Sanctuary AI}} & Phoenix Gen 7 & 2018 & Vancouver, BC, Canada & Pilot/Trials & Retail, Manufacturing & Proprietary (Carbon AI) & Embodied AI, human-like hands & --- \\
\rowcolor{unitree_light!30}
\textcolor{unitree_primary}{\textbf{Tesla}} & Optimus Gen 2 & 2003 & Austin, TX, USA & Development & Manufacturing, Logistics & Proprietary & Tesla-designed actuators, 11 DoF hands & \$20k--\$30k (target) \\
\textcolor{unitree_primary}{\textbf{Toyota Research Institute}} & T-HR3 & 2016 & Los Altos, CA, USA & Research & Remote operation, Service & Proprietary research stack & Master--slave teleoperation, torque servo & Research only \\
\rowcolor{unitree_light!30}
\textcolor{unitree_primary}{\textbf{UBTECH Robotics}} & Walker X & 2012 & Shenzhen, China & Commercial & Service, Entertainment & Proprietary & 41 DoF, visual navigation & --- \\
\textcolor{unitree_primary}{\textbf{Unitree Robotics}} & G1, H1 & 2016 & Hangzhou, China & Commercial & Research, Industrial & SDK (C++/Python); ROS support & Affordable platforms, 23 DoF & \$16,000+ \\
\rowcolor{unitree_light!30}
\textcolor{unitree_primary}{\textbf{Xiaomi}} & CyberOne & 2010 & Beijing, China & Prototype & Consumer, Service & Proprietary & Emotion recognition, 21 DoF & --- \\

\end{longtable}
\end{landscape}


\renewcommand{\UrlFont}{\small\ttfamily}
\bibliographystyle{unsrtnat}
\bibliography{bibliography}

\begin{thebibliography}{17}
\providecommand{\natexlab}[1]{#1}
\providecommand{\url}[1]{\texttt{#1}}
\expandafter\ifx\csname urlstyle\endcsname\relax
  \providecommand{\doi}[1]{doi: #1}\else
  \providecommand{\doi}{doi: \begingroup \urlstyle{rm}\Url}\fi

\bibitem[Mayoral-Vilches(2025{\natexlab{a}})]{mayoral2025offensive}
V{\'\i}ctor Mayoral-Vilches.
\newblock Offensive robot cybersecurity.
\newblock \emph{arXiv preprint arXiv:2506.15343}, 2025{\natexlab{a}}.

\bibitem[Quigley et~al.(2009)Quigley, Conley, Gerkey, Faust, Foote, Leibs,
  Wheeler, Ng, et~al.]{quigley2009ros}
Morgan Quigley, Ken Conley, Brian Gerkey, Josh Faust, Tully Foote, Jeremy
  Leibs, Rob Wheeler, Andrew~Y Ng, et~al.
\newblock Ros: an open-source robot operating system.
\newblock In \emph{ICRA workshop on open source software}, volume~3, page~5.
  Kobe, 2009.

\bibitem[Mayoral-Vilches et~al.(2020)]{mayoral2020robot}
V{\'\i}ctor Mayoral-Vilches et~al.
\newblock Robot cybersecurity, a review.
\newblock \emph{International Journal of Cyber Forensics and Advanced Threat
  Investigations}, 2020.
\newblock URL \url{https://conceptechint.net/index.php/CFATI/article/view/41}.

\bibitem[Mayoral-Vilches et~al.(2022)Mayoral-Vilches, Garc{\'\i}a-Maestro,
  Towers, and Gil-Uriarte]{mayoral2022alurity}
V{\'\i}ctor Mayoral-Vilches, Nuria Garc{\'\i}a-Maestro, Michael Towers, and
  Endika Gil-Uriarte.
\newblock Alurity: A systematic review on the security of robotic and
  autonomous systems.
\newblock \emph{arXiv preprint arXiv:2203.13874}, 2022.

\bibitem[Surve et~al.(2024)Surve, Shabtai, and Elovici]{surve2024sok}
Priyanka~Prakash Surve, Asaf Shabtai, and Yuval Elovici.
\newblock Sok: Cybersecurity assessment of humanoid ecosystem, 2024.
\newblock URL \url{https://arxiv.org/abs/2508.17481}.

\bibitem[{Investing.com}(2025)]{investing2025tesla}
{Investing.com}.
\newblock Musk: Tesla to build 5000 optimus robots in 2025, tells staff, March
  2025.
\newblock URL
  \url{https://www.investing.com/news/stock-market-news/musk-tesla-to-build-5000-optimus-robots-in-2025-urges-staff-not-to-sell-shares-3941167}.
\newblock Accessed: 2025-09-14.

\bibitem[{BMW Group}(2024)]{bmw2024figure}
{BMW Group}.
\newblock Figure announces commercial agreement with bmw manufacturing to bring
  general purpose robots into automotive production, January 2024.
\newblock URL
  \url{https://www.prnewswire.com/news-releases/figure-announces-commercial-agreement-with-bmw-manufacturing-to-bring-general-purpose-robots-into-automotive-production-302036263.html}.
\newblock Accessed: 2025-09-14.

\bibitem[{Goldman Sachs Research}(2024)]{goldmansachs2024humanoid}
{Goldman Sachs Research}.
\newblock Global automation: Humanoid robot - the ai accelerant.
\newblock Technical report, Goldman Sachs, 2024.
\newblock URL
  \url{https://www.goldmansachs.com/insights/articles/the-global-market-for-robots-could-reach-38-billion-by-2035}.
\newblock Projects \$38 billion humanoid robot market by 2035.

\bibitem[{IDTechEx}(2025)]{idtechex2025humanoid}
{IDTechEx}.
\newblock Humanoid robots 2025-2035: Technologies, markets and opportunities.
\newblock Technical report, IDTechEx Research, 2025.
\newblock URL
  \url{https://www.idtechex.com/en/research-report/humanoid-robots/1093}.
\newblock Forecasts \$30 billion market by 2035.

\bibitem[{IEEE Spectrum}(2024)]{spectrum2024humanoid}
{IEEE Spectrum}.
\newblock Humanoid robots: The scaling challenge.
\newblock \emph{IEEE Spectrum}, 2024.
\newblock URL \url{https://spectrum.ieee.org/humanoid-robot-scaling}.
\newblock Critical analysis of humanoid robot market projections.

\bibitem[Kirschgens et~al.(2018)Kirschgens, Ugarte, Uriarte, Rosas, and
  Vilches]{kirschgens2018robot}
Laura~Alzola Kirschgens, Irati~Zamalloa Ugarte, Endika~Gil Uriarte, Aday~Muniz
  Rosas, and V{\'\i}ctor~Mayoral Vilches.
\newblock Robot hazards: from safety to security.
\newblock \emph{arXiv preprint arXiv:1806.06681}, 2018.

\bibitem[Shin et~al.(2017)Shin, Kim, Kwon, and Kim]{shin2017illusion}
Hocheol Shin, Dohyun Kim, Yujin Kwon, and Yongdae Kim.
\newblock Illusion and dazzle: Adversarial optical channel exploits against
  lidars for automotive applications.
\newblock In \emph{International Conference on Cryptographic Hardware and
  Embedded Systems}, pages 445--467. Springer, 2017.

\bibitem[Petit et~al.(2015)Petit, Stottelaar, Feiri, and
  Kargl]{petit2015remote}
Jonathan Petit, Bas Stottelaar, Michael Feiri, and Frank Kargl.
\newblock Remote attacks on automated vehicles sensors: Experiments on camera
  and lidar.
\newblock In \emph{Black Hat Europe}, volume~11, page 2015, 2015.

\bibitem[Mayoral-Vilches(2025{\natexlab{b}})]{mayoralvilches2025cybersecurityaidangerousgap}
Víctor Mayoral-Vilches.
\newblock Cybersecurity ai: The dangerous gap between automation and autonomy,
  2025{\natexlab{b}}.
\newblock URL \url{https://arxiv.org/abs/2506.23592}.

\bibitem[Quarta et~al.(2017)Quarta, Pogliani, Polino, Maggi, Zanero, and
  Zanchettin]{quarta2017experimental}
Davide Quarta, Marcello Pogliani, Mario Polino, Federico Maggi, Stefano Zanero,
  and Andrea~Maria Zanchettin.
\newblock An experimental security analysis of an industrial robot controller.
\newblock In \emph{2017 IEEE Symposium on Security and Privacy (SP)}, pages
  268--286. IEEE, 2017.

\bibitem[Son et~al.(2015)Son, Shin, Kim, Park, Noh, Choi, Choi, and
  Kim]{son2015rocking}
Yunmok Son, Hocheol Shin, Dongkwan Kim, Youngseok Park, Juhwan Noh, Kibum Choi,
  Jungwoo Choi, and Yongdae Kim.
\newblock Rocking drones with intentional sound noise on gyroscopic sensors.
\newblock In \emph{24th USENIX Security Symposium}, pages 881--896, 2015.

\bibitem[Mayoral-Vilches et~al.(2025)Mayoral-Vilches, Navarrete-Lozano,
  Sanz-Gómez, Espejo, Crespo-Álvarez, Oca-Gonzalez, Balassone, Glera-Picón,
  Ayucar-Carbajo, Ruiz-Alcalde, Rass, Pinzger, and
  Gil-Uriarte]{aliasrobotics2025cai}
Víctor Mayoral-Vilches, Luis~Javier Navarrete-Lozano, María Sanz-Gómez,
  Lidia~Salas Espejo, Martiño Crespo-Álvarez, Francisco Oca-Gonzalez,
  Francesco Balassone, Alfonso Glera-Picón, Unai Ayucar-Carbajo, Jon~Ander
  Ruiz-Alcalde, Stefan Rass, Martin Pinzger, and Endika Gil-Uriarte.
\newblock Cai: An open, bug bounty-ready cybersecurity ai, 2025.
\newblock URL \url{https://arxiv.org/abs/2504.06017}.

\end{thebibliography}

\end{document}